\documentstyle[epsfig]{mn}

\input{epsf.sty}

\def\mpc {h^{-1}~{\rm{Mpc}}}
\def\kpc {h^{-1} {\rm{kpc}}}
\def\msun {{\rm M}_{\odot}}

\begin{document}

\title
[The 2QZ - X. Lensing by galaxy groups]
{
The 2dF QSO Redshift Survey - X.  Lensing of Background QSOs by Galaxy Groups.
}

\author[A.~D. Myers et~al. ]
{
A.~D. Myers$^1$, P.~J. Outram$^1$, T. Shanks$^1$, B.~J. Boyle$^{2}$, S. M. Croom$^{2}$, 
\newauthor   N.~S. Loaring$^{3}$, L. Miller$^{3}$ \& R.~J. Smith$^{4}$
\\
1 Department of Physics, Science Laboratories, South Road, Durham, DH1 3LE, U.K.
\\
2 Anglo-Australian Observatory, PO Box 296, Epping, NSW 2121, Australia\\
3 Department of Physics, Oxford University, Keble Road, Oxford, OX1 3RH, U.K.\\
4 Liverpool John Moores University, Twelve Quays House, Egerton Wharf, Birkenhead, CH41 1LD, U.K. \\
}

\maketitle 
 
\begin{abstract}

We cross-correlate QSOs from the 2dF
QSO Redshift Survey with groups of galaxies.  In the southern region of
the 2dF we utilise galaxies from the APM Survey.  In the northern strip,
galaxies are taken from the recent Sloan Digital Sky Survey Early Data
Release.  Both galaxy samples are limited to a depth $B < 20.5$.  We use
an objective clustering algorithm to detect groups in these galaxy catalogues.

A $3\sigma$ anti-correlation is observed between 2dF QSOs and
galaxy groups, confirming the effect found by Boyle, Fong \& Shanks in
an independent dataset.  This paucity of faint QSOs around
groups cannot be readily attributed to a selection effect and is
not due to restrictions on the placement of 2dF fibres.  By
observing the colours of QSOs on the scales of the anti-correlation, we 
limit the influence intervening dust in galaxy groups can have on
background QSO flux, finding a maximum reddening on the scale of the
anti-correlation of $E(b_j-r) \leq 0.012$ at the 95~per~cent level.  The small amount of dust inferred
from the QSO colours would be  insufficient to account for the
anti-correlation, supporting the suggestion by Croom \& Shanks that the
signal is likely caused by weak gravitational lensing.  The
possibility remains that tailored dust models involving grey dust, heavy 
patches of dust, or a combination of dust and lensing, could explain the anti-correlation. 

Under the assumption that the signal is caused by lensing rather than
dust, we measure the average velocity dispersion of a
Singular Isothermal Sphere that would cause the anti-correlation,
finding $\sigma \sim 1150 \rm{~km~s^{-1}}$.  Simple simulations reject
$\sigma \sim 600 \rm{~km~s^{-1}}$ at the 5~per~cent significance level.
We also suppose the foreground mass
distribution consists of dark matter haloes with an NFW profile and
measure the typical mass within $1.5~\mpc$ of the halo centre as
$M_{1.5} = 1.2\pm{0.9}  \times 10^{15}~h^{-1}~\msun$.  

Regardless of whether we utilise a Singular Isothermal Sphere or NFW
dark matter profile, our simple lensing model favours more mass in groups of
galaxies than accounted for in a universe with density parameter 
$\Omega_m = 0.3$.   Detailed simulations and galaxy group redshift
information will significantly reduce the current systematic
uncertainties in these $\Omega_m$ estimates.
Reducing the remaining statistical uncertainty in this result 
will require larger QSO and galaxy group surveys.  

\bigskip
\end{abstract}

\begin{keywords}
{surveys - quasars, quasars: general, large-scale structure of
Universe, cosmology: observations, gravitational lensing}
\end{keywords}

\section{Introduction}

Claims of QSO-foreground galaxy associations are now more usually
interpreted in terms of gravitational lensing rather than evidence for
non-cosmological redshifts (Bukhmastova 2001; Benitez, Sanz \&
Martinez-Gonzalez 2001).  Observationally, however, the situation has been
complex, with a variety of effects recorded in the literature.  For example
Williams \& Irwin (1998) detected a positive correlation between
high-redshift LBQS QSOs and APM galaxies on degree scales, yet
Martinez et al. (1999) find no strong quasar-galaxy angular correlation 
on similar scales.   Ferreras, Benitez \& Martinez-Gonzalez (1997)
found a large anti-correlation between galaxies and optically selected
QSOs near the NGP, and suggested a difficulty in selecting QSOs in
densely populated areas.  Samples of QSOs utilised in these and earlier 
papers frequently suffered from either inhomogeneity or a dearth of data
(see, e.g., Norman \& Williams 2000, for a review).  Recently, Norman \& Impey (2001) have
found a significant positive correlation between a homogeneous sample of
90 radio-bright QSOs (with median optical magnitude $V \sim 18.5$) and
galaxies lying outside of rich clusters.  The amplitude of this angular
correlation was consistent with a lensing explanation.  The completed
2dF QSO Redshift Survey contains a UVX-selected homogeneous sample of
around 23,000 QSOs. Contemporary models suggest that statistical lensing
should become a stark cosmological effect in such vast surveys \cite{Men02}.

Following Shanks et al. (1983), Boyle, Fong \& Shanks (1988, henceforth BFS88)  quoted a
significant anti-correlation between galaxies in objectively selected
clusters and faint UVX objects.  Initially, BFS88 interpreted the anti-correlation as an
effect caused by a small amount of dust in foreground galaxy groups obscuring background QSOs.
Ferguson (1993) and Maoz (1995) generally restricted reddening in
clusters and rich groups, results that marginally suggested
insufficient dust to induce the observed lack of QSOs around
galaxy groups.  This prompted Croom \& Shanks (1999) to recast the anti-correlation signal in terms of
statistical lensing.  Rodrigues-Williams \& Hogan (1994) first
discussed the BFS88 anti-correlation result as a possible effect of
gravitational lensing in a paper confirming that bright QSOs are positively
correlated with galaxy clusters.  Significant positive correlations between galaxies and bright 
QSOs, which cannot be explained away by dust in galaxies, continue to be
detected (Williams \& Irwin 1998; Norman \& Williams 2000; Norman \&
Impey 2001; Gaztanaga 2002).

Gravitational lensing can satisfactorily explain QSO-galaxy
associations.  Interim mass lenses the area behind it,  influencing a 
sample of distant objects in two related ways.  Firstly, sources are
magnified.  Secondly, the apparent sky density of sources of given
intrinsic luminosity drops.  If the
number-magnitude relation of the considered sample is steep, the first 
effect dominates and we observe the more numerous, fainter QSO population.  If the
number-magnitude slope is shallow, the second effect prevails and we
see fewer QSOs, as the area that they populate dilutes.
Ultimately, this magnification bias will increase the correlation of QSOs and
galaxies if we consider a sample near an intrinsically steep part of the
QSO number-magnitude relation, and it will induce a paucity of QSOs
around foreground mass where QSO number counts flatten \cite{Wu94}.
The optical QSO number counts are steep for bright QSOs and flatten
significantly at faint magnitudes.  Observationally, then, we would
expect both positive and negative correlations between QSOs and galaxies,
depending on the apparent luminosity of the QSO sample.

Gravitational lensing techniques have fast become popular methods of
tracing the mass distributions in our Universe; see Wambsganss (1996)
for a review.  The attraction is obvious - no assumptions need be
made about the virialisation of the observed matter, or about how the
luminous matter traces the underlying distribution of mass.
Reconstructing individual cluster masses from the shear and magnification of
background galaxies is a well-established cosmological tool  (Squires
\& Kaiser 1996 and references therein) and measurements of the
statistical influence of poor groups on the background galaxy
population can probe the distribution of matter in the field
\cite{Hoe01}.   Broadhurst, Taylor and Peacock (1995) have suggested
using magnification bias to measure the effect dark matter in a cluster 
has in lensing the overall distribution of background galaxies.
Studying the influence clusters exert on background QSOs is,
perhaps, an even better approach; although QSO samples are significantly smaller than
galaxy samples, background QSOs are more readily distinguished from cluster
members than background galaxies.

Questions remain about the anti-correlation between faint UVX objects
and galaxies in groups detected by BFS88.  Is
the result reproduced in a different sample?  Is it affected by dust in
galaxy groups?  Is it a selection effect or a systematic?  Is statistical 
lensing a viable explanation given the large amplitude of the anti-correlation?
In this paper, we address these questions by measuring the two-point
correlation function between objects in the 2dF QSO Redshift Survey
\cite{Cro01} and galaxy groups determined
from the APM catalogue  \cite{Mad90a} and from the Sloan Digital Sky
Survey Early Data Release \cite{Sto02}.   The 2dF QSO redshift survey contains a
large, homogeneous, objectively determined sample of objects.  The QSOs are spectroscopically
confirmed, meaning contamination is extremely low. The 2dF survey also
measures colour, allowing strict limits to be placed on the effect of
dust in galaxies on background QSOs.  Finally, in modelling our lensing signal,
we consider a different cluster profile to BFS88 and adapt our
analysis so that the model more fairly represents the data. 

This paper primarily deals with the cross-correlation of QSOs versus
foreground galaxy groups and what it indicates.  In the following section we
discuss the 2dF QSO Redshift Survey and derive
necessary results from it.  In Section 3 we outline the galaxy samples we
cross-correlate with QSOs and the objective method by
which we derive galaxy groups.  In Section 4 we display the
cross-correlation analysis, discussing the possibilities that its form is 
attributable to selection effects or dust.  In Section 5 we interpret
our results as indicative of statistical gravitational lensing and
discuss the implications of such an interpretation.  Section 6 presents 
our conclusions.

\section{The 2dF QSO Redshift Survey}

The 2dF QSO Redshift Survey (henceforth 2QZ) is named for the 2-degree
Field multi-object spectrograph being utilised at the AAT to survey
740~deg$^2$ of sky.  The 2QZ patch is contained within the 2dF
Galaxy Redshift Survey \cite{Col98}, and comprises $75\degr$ in ascension and
$5\degr$ in declination in the region of both the North and South Galactic
Caps.  In the North, the strip is bounded by $-2\fdg5$ and $2\fdg5$
Declination, and $9^{\rm{h}}~50^{\rm{m}}$ and $14^{\rm{h}}~50^{\rm{m}}$
Right Ascension. The southern strip extends from $-32\fdg5$ to
$-27\fdg5$ Declination and from $21^{\rm{h}}~40^{\rm{m}}$ to
$3^{\rm{h}}~15^{\rm{m}}$ Right Ascension.  Missing area can be attributed to
holes containing cuts for bright stars.  

The QSO target sample, the input catalogue \cite{Smi97}, was selected
from colour cuts in the $(u-b_j)$ against $(b_j -r)$ plane of APM
scanned UKST data to limiting magnitude $18.25 \leq b_j \leq 20.85$ ($18.4 \la b \la
21$).  At the bright end, the survey has been extended to $b_j > 16$
using the 6-degree Field spectrograph.  Spectra of each object in the input catalogue were taken, to
determine what percentage are genuinely QSOs.  2QZ spectroscopic
observations have been carried out since 1997, with a 
projected identification of 25,000 QSOs.  Each 2dF QSO survey field was
observed by 125 optical fibres for around an hour. Many of the
remaining 275 fibres concurrently observed galaxies for the Galaxy
Redshift Survey. The minimum fibre separation was of the order of 20~arcsec on
the sky.  The spectra cover the wavelength range 3700-8000\AA. The
process yielded an average signal-to-noise ratio of around 5  in the central
wavelength range of the faintest sources, allowing categorical
spectroscopic identification of objects in all but the poorest seeing. For
further general information on the technicalities of the survey,
consider Boyle et al. (2000) or Croom et al. (2001).

The 2QZ is now essentially complete.  Spectra have been measured for nearly
45,000 objects, of which about 12,000 are categorically identified as
stars (or White Dwarfs) and about 23,000 as QSOs.  The catalogue also
contains a number of Narrow Emission Line Galaxies (NELGs), BL Lac objects and
unidentified, or low signal-to-noise, spectra.  The equatorial strip
near the NGC (North Galactic Cap) contains nearly 10,500 of the QSOs.
The strip in the SGC (South Galactic Cap) contains over 12,800 QSOs.

\subsection{Number counts by magnitude}

The expected strength of lensing-induced correlations between galaxies and a magnitude-limited
sample of QSOs depends on the slope of the integrated number-magnitude
counts, $\beta$, fainter than the QSO sample's limit \cite{Nar89}.  We thus need
to estimate this slope to interpret our results.  As we do not
have information fainter than the limit of the 2QZ, we
fit models to the 2QZ number-counts brighter than the $b_j =
20.85$ limit and extrapolate the counts to fainter magnitudes.
Such models are better constrained by fitting them to the differential
counts and integrating them.  

In Figure 1 we present the differential number counts by magnitude of
the completed 2QZ, which have been extensively corrected for
incompleteness and dust (see Boyle et al., in preparation) and averaged
over both hemispheres.  Also plotted are points from the 6dF QSO Survey
(6QZ).  When determining the number-count relation, the sample is restricted to the redshift range
$0.3<z<2.2$, the range for which the 2QZ is designed to be photometrically
complete \cite{Cro01}.  Error bars are Poisson.  

\begin{figure} 
\begin{centering}
\begin{tabular}{c}
{\epsfxsize=8truecm \epsfysize=8truecm \epsfbox[35 170 550 675]{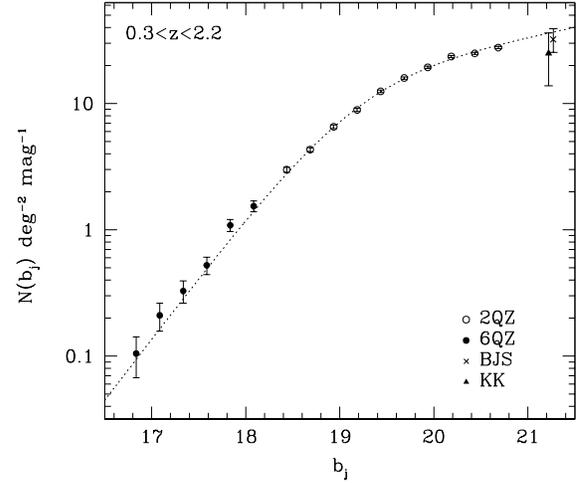}}
\end{tabular}
\caption
{The differential number counts of the
  2dF QSO Redshift Survey.  The points are QSO number counts in 0.2~mag
  bins, with Poisson errors. The dashed line ia a smoothed power law fit to
  the data.  Brighter data points are from the 6dF QSO Redshift Survey.
  Also displayed are the faint data from Boyle, Jones \& Shanks (1991) and Koo \& Kron (1988).}
\label{fig:nmagdiff.ps}
\end{centering}
\end{figure}

\begin{figure} 
\begin{centering}
\begin{tabular}{c}
{\epsfxsize=8truecm \epsfysize=8truecm \epsfbox[35 170 550 675]{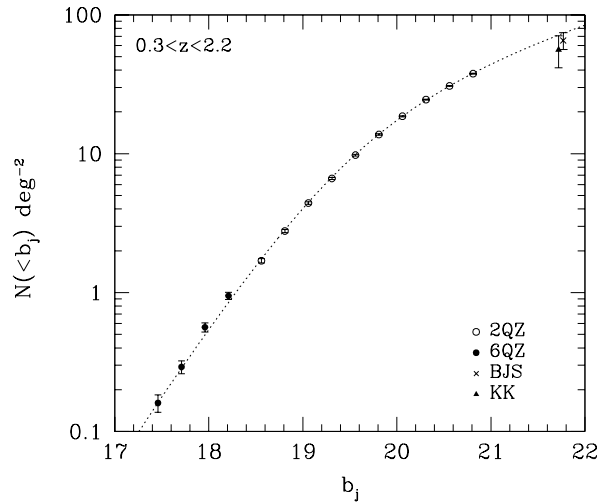}}
\end{tabular}
\caption
{The $N(<m)$ relationship, or integrated number counts, for the
  2dF QSO Redshift Survey.  The points are QSO number counts in 0.2~mag
  bins, with Poisson errors. The line is a smoothed power law fit to
  the differential number counts.  Brighter data points are from the
  6dF QSO Redshift Survey. Also displayed are the faint data 
  from Boyle, Jones \& Shanks (1991) and Koo \& Kron (1988), which have 
  been offset slightly to prevent the points from merging.}

\label{fig:Nmag.ps}
\end{centering}
\end{figure}

\begin{figure} 
\begin{centering}
\begin{tabular}{c}
{\epsfxsize=8truecm \epsfysize=8truecm \epsfbox[35 170 550 675]{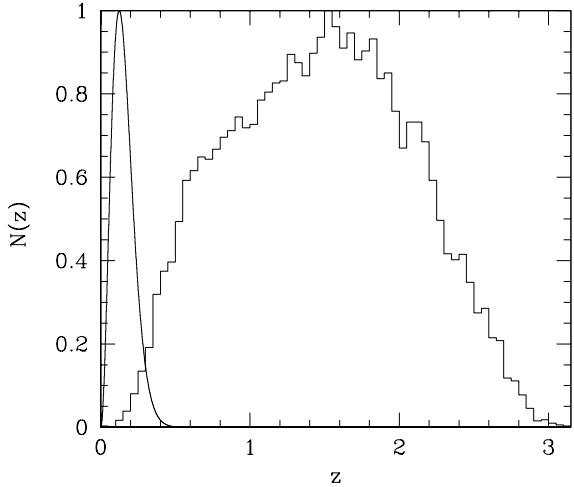}}
\end{tabular}
\caption
{The N(z) relationship for the 2dF QSO Redshift Survey.  The wider
  histogram represents QSO number counts in 0.05 redshift bins.  The
  narrow histogram is an analytic model of the galaxy distribution at
  the B = 20.5 limit of the galaxy samples used in this paper. The
  distributions are normalised to reach a maximum at unity.}

\label{fig:Nz.ps}
\end{centering}
\end{figure}

The dashed line is a smoothed power law model, where the differential counts
are expressed in the form

\begin{eqnarray}
\frac{\rm d \it N}{\rm d \it m} = \frac{N_0}{10^{-\alpha_{\rm{d}}(m-m_0)} + 10^{-\beta_{\rm{d}}(m-m_0)}}
\end{eqnarray}

The best-fit model has a bright-end slope of $\alpha_{\rm{d}} = 0.98$, a knee at $m_0 
= 19.1$ and a faint-end slope of $\beta_{\rm{d}} = 0.15$.  This model is consistent
with faint data from Boyle, Jones \& Shanks (1991)
and Koo \& Kron (1988), which are also marked in Figure 1.  Data from
the literature have undergone a zeropoint correction of $b_j = b -
0.15$ and a dust correction of about 0.1~mag.  

In Figure 2, the models are integrated and displayed against the
integrated number-magnitude counts.  The best-fit model to the
differential counts has an average integrated
faint-end slope of $\beta = 0.29$.  Although
the model is well constrained by the data ($\sigma = \pm 0.015$) the
many incompleteness corrections to the faint-end data (again, see Boyle 
et al., in preparation) mean the $1\sigma$ error may be as high as $\pm
0.05$.  When BFS88 modelled the magnitude
distribution of QSOs as a broken power law, they determined a B-band
faint end slope of $0.32-0.33$. In an extensive review, Hartwick \& Schade (1990)
subsequently determined a faint-end slope of $0.31$.  Our average slope 
is thus consistent with these earlier authors.

\subsection{Number counts by redshift}

When making model predictions to interpret our results, we shall need to assign both QSOs and
galaxies a redshift, in order to estimate the average QSO-galaxy
separation (for a given cosmology).  We randomly select QSO redshifts from
the number-redshift distribution of the 2QZ and galaxy
redshifts from the analytic distribution of Baugh \& Efstathiou
(1993).  The average QSO-galaxy separation is then estimated as the mean
separation from thousands of random pairs of redshifts.

In Figure 3 we display the number counts by redshift
in the 2QZ as a histogram.  Also marked is an analytic model for a
galaxy redshift distribution \cite{Bau93} integrated to $B = 20.5$, the
galaxy sample limit we typically consider in this paper.  Both
distributions have been normalised to peak at 1.  Note that for redshifts greater
than 0.4, less than 0.4~per~cent of the projected galaxy distribution
overlaps the QSO distribution.  

\section{APM and SDSS clusters}

\subsection{The catalogues}

Our correlation analysis in subsequent sections relies on galaxies from two generations
of surveys, the APM Galaxy Survey and the Early Data Release (EDR) of the
Sloan Digital Sky Survey (SDSS).  

\begin{figure*} 
\begin{centering}
\begin{tabular}{c}
\centerline{\epsfig{figure=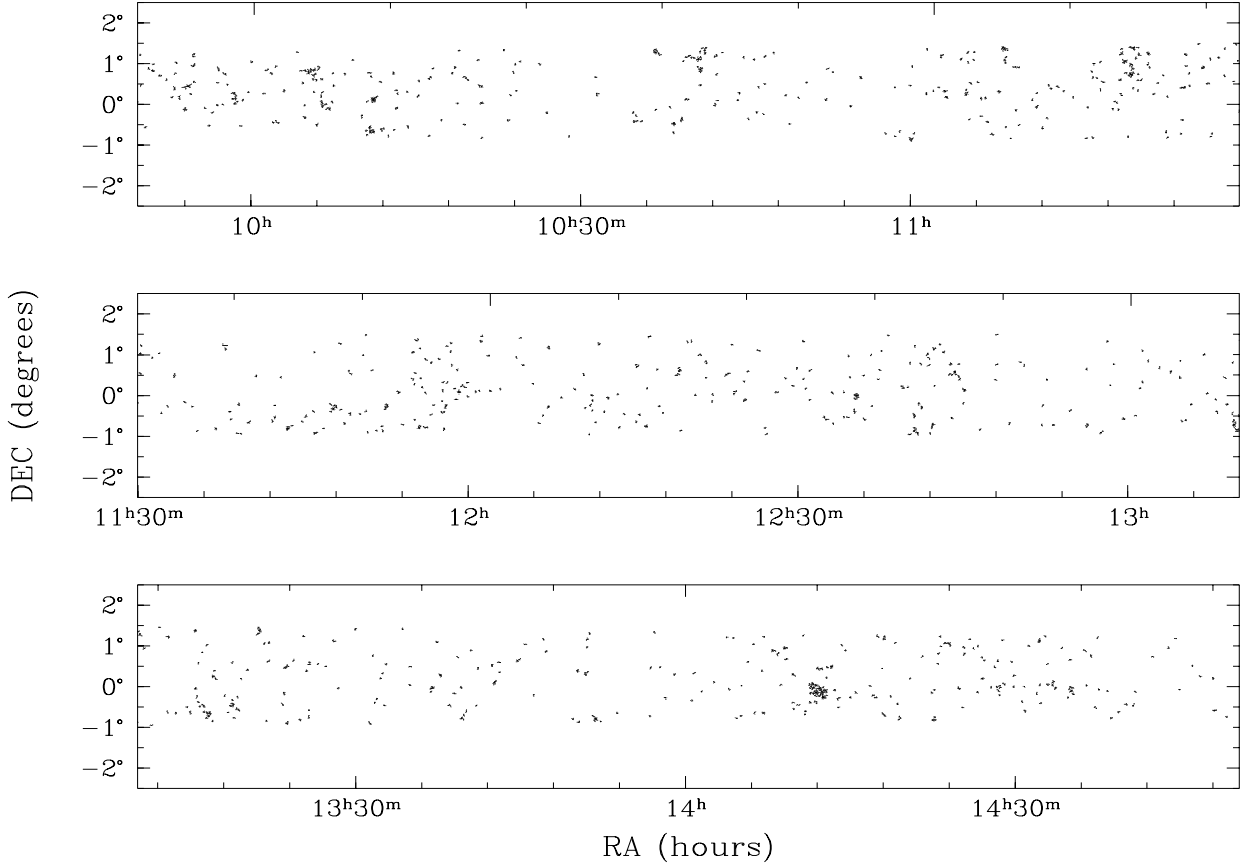,width=5in,height=7in,angle=-90}}
\end{tabular}
 \caption{Objectively defined galaxy groups in the Sloan Digital Sky
   Survey Early Data Release.  Each point is a galaxy in a region
   sufficiently dense that it meets clustering criteria outlined in the 
   text. The axes correspond to the limits of the 2dF QSO Redshift
   Survey.  The coordinate system has been transformed to B1950.}
\label{fig:SDSS groups}
\end{centering}
\end{figure*}

The initial APM survey \cite{Mad90a} was derived from Automated Plate
Measuring scans of 185 photographic UKST plates and covered about 10~per~cent of
the entire sky around the SGP.  
The original APM region was bounded between a Right Ascension of roughly 
$21^{\rm{h}}$ to $5^{\rm{h}}$, with Declination from $-72\degr$ to
$-18\degr$. Images were detected to a B-band magnitude of 21.5, 
allowing galaxies to be definitively identified down to $B < 20.5$.
The photometry was extensively aligned using overlapping plates
\cite{Mad90b}.   The APM study was later extended and provided
the input catalogues for the 2dF surveys.  The original APM completely covers the
southern 2QZ strip.

\begin{figure*} 
\begin{centering}
\begin{tabular}{c}
\centerline{\epsfig{figure=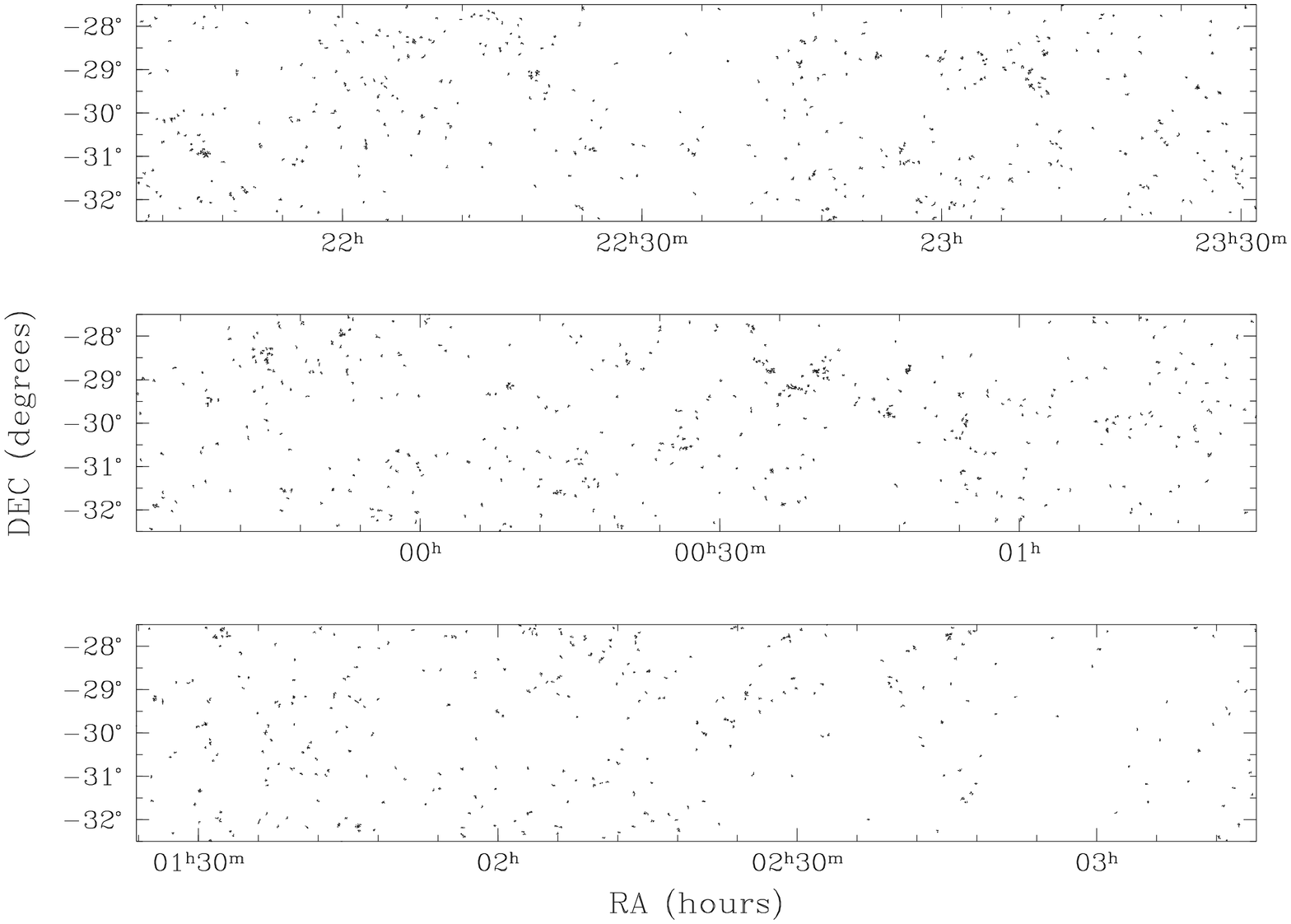,width=5in,height=7in,angle=-90}}
\end{tabular}
 \caption{Objectively defined galaxy groups in the Southern APM
   Survey.  Each point is a galaxy in a region
   sufficiently dense that it meets clustering criteria outlined in the 
   text. The axes correspond to the limits of the 2dF QSO Redshift
   Survey.  The coordinate system is B1950.}
\label{fig:APM groups}
\end{centering}
\end{figure*}

The SDSS (http://www.sdss.org) is imaging the northern sky in
five bands designed for CCD photometry ($u', g', r', i'$ and $z'$). 
The survey will trace an ellipse centred on $12^{\rm{h}}~20^{\rm{m}}$
Right Ascension and $32.8\degr$ Declination, roughly
extending from $7^{\rm{h}}~6^{\rm{m}}$ to $17^{\rm{h}}~34^{\rm{m}}$ in
Right Ascension and $\pm 55\degr$ of declination.   The SDSS should be
complete to $g' \sim 23.3$ and $r' \sim 23.1$ \cite{Yor00}, about
equivalent to $b \sim 23.5$ using a
typical colour transformation \cite{Yas01}.  The Early Data Release \cite{Sto02} contains about
the first $460~\deg^{2}$ of the SDSS.  Conveniently, one strip,
extending from  $9^{\rm{h}}~40^{\rm{m}}$ to $15^{\rm{h}}~44^{\rm{m}}$ in
Right Ascension and $-1\degr$  to $1.5\degr$ Declination, is contained 
within the northern 2QZ strip.

\subsection{Objective group catalogues}

We wish to correlate the 2QZ sample with groups of galaxies rather than 
the field as any correlation attributed to lensing will be stronger for more massive
foreground structures.  A catalogue of objectively identified groups in the Southern APM is
available in the literature \cite{Dal97} but comparatively little has
been available in the region of the northern 2QZ since the publication
of the ACO catalogue \cite{Abe89}.  Additionally, it is useful to have listings of the 
position of each galaxy within each cluster, not merely the positions
of cluster centres.  Following BFS88 we turn to the clustering routine of
Turner \& Gott (1976) to objectively identify clusters of galaxies in the 
APM survey and SDSS EDR.

To determine the boundaries for cluster membership in a galaxy sample,
we assign an overdensity, $\delta$, a factor by which we wish our
group density to exceed the mean surface density across the entire
region we consider ($\bar{\sigma}$). We then calculate the largest
possible circle of radius $\theta_c$ such that

\begin{eqnarray}
\sigma(\theta<\theta_c)>\delta\bar{\sigma}
\end{eqnarray}

where $\sigma$ is the surface density of galaxies within the (circular) 
region centred on any particular galaxy in our sample and enclosed by
an angular radius $\theta$.  Over the miniscule angles typically considered,
$\sigma$ can be expressed as

\begin{eqnarray}
\sigma = \frac{N(\leq\theta)}{2\rm{\pi}(1-\cos\theta)} \approx \frac{N(\leq\theta)}{\rm{\pi}\theta^2}
\end{eqnarray}

where $N(\leq\theta)$ is the total number of galaxies within an angle
$\theta$ of the particular galaxy we are considering, including the
particular galaxy itself. The critical angular radius $\theta_c$
is determined for each individual galaxy in the sample.  A group is
defined as all galaxies that have overlapping critical radii.

Two classification questions remain;  how overdense are groups (what
value should $\delta$ take)? and what size must groups attain before we 
call them a group (what is the minimum number of galaxies, $N_{min}$, in a
group)?  We take the values of $\delta = 8$ and $N_{min} =
7$ chosen by Stevenson, Fong \& Shanks (1988) in a similar analysis,
and used by BFS88.
The choice of the overdensity parameter was originally suggested by Turner \& Gott (1976)
and weighs the possibility of losing poor clusters at high $\delta$
values against false grouping of galaxies in the field at low $\delta$.
The choice of $N = 7$ as a minimum group size reduces the likelihood of
chance alignments of galaxies at different redshifts being grouped.
The parameters reflect a distribution of groups that may have been
selected by eye \cite{Ste88}.

Figures 4 and 5 show the groups determined from the SDSS EDR data in the 
northern strip and from APM data in the southern strip.  In both cases, 
the galaxy sample was limited to $B < 20.5$, the limit of the APM.  The
SDSS EDR data was transformed from the SDSS photometric system using
the colour transformation of Yasuda et al. (2001).  The SDSS EDR
coordinate system has been transformed from J2000 to B1950, the system
we shall use throughout this paper.  The axes in both figures correspond
to the limits of the 2QZ.  
A great deal of the structure produced by Dalton et al. (1997) is reproduced well in our
southern group catalogue, even though Dalton used different clustering
criteria.  30~per~cent of Dalton's cluster centres
within the 2QZ boundary lie within 2~arcmin of our group centres, improving to 75~per~cent of
Dalton's cluster centres lying within 5~arcmin of our own.  90~per~cent 
of group centres match up within a separation of 7.7~arcmin, the
mean cluster radius of Dalton's groups within the 2QZ SGC region.  Many of
the larger groups automatically picked out by our technique also correspond
to previously identified rich clusters \cite{Abe58}, for instance,
the large cluster around $14^{\rm{h}}~12^{\rm{m}}$ Right Ascension in
Figure 4 corresponds to Abell 1882 (Richness Class III, $z \sim 0.137$).
Abell assigned 166 galaxies to Abell 1882, we determine 153 members.
Our voids are also in good agreement with the literature.  The sparse
region we pick out in the Southern APM from Right Ascension
$2^{\rm{h}}~30^{\rm{m}}$ to $3^{\rm{h}}$ appears to be real.  In this
region, Dalton (1997) finds no clusters over the Declination range
$-32\fdg5$ to $-28\degr$ and Abell finds a single (Richness Class 0)
cluster in the range $-32\fdg5$ to $-29\fdg5$.  The void is not an
effect of intervening dust obscuring galaxies out of the APM - the region of the SGC strip from
$23^{\rm{h}}$ to $23^{\rm{h}}~30^{\rm{m}}$ has more dust across it
(according to the maps of Schlegel, Finkbeiner \& Davis 1998) and a 
denser population of groups.  In any case, we note that similar voids
are picked out at, say, $22^{\rm{h}}~30^{\rm{m}}$ in the Southern APM, or 
$10^{\rm{h}}~30^{\rm{m}}$ in the SDSS EDR.

\section{Cross-correlation analysis}

We now turn our attention to cross-correlations of galaxy groups with
2QZ QSOs.

\subsection{Method}

The two-point angular correlation function $\omega(\theta)$  measures
the probability (d$\it{P}$) of finding pairs of sources of mean number
density $\bar{n}$ within solid angle d$\it{\Omega}$ separated by angle $\theta$

\begin{eqnarray}
  \rm{d} \it{P} = \it{\bar{n}\rm_1 \it \bar{n} \rm _2}(\rm 1+\it  \omega(\theta))\rm d \it \Omega \rm _1\rm d \it \Omega \rm _2
\end{eqnarray}

The correlation function then, measures the relative amount of associated
structure in pairs of sources for various angular separations.  If
there is some association the correlation function will be positive.
If there is some avoidance, the correlation function will be negative.
If there is no distributed pattern whatsoever, as we might expect from, 
say, stars versus extragalactic objects, the correlation function will be zero. 
Throughout this analysis, we estimate the two-point correlation
function using an equation of the form proposed by Peebles (1980)

\begin{eqnarray}
\omega(\theta) = \frac{DD(\theta)\bar{n} \rm{_R}}{DR(\theta)\bar{n} \rm{_D}} - 1
\end{eqnarray}

where D refers to a data point (either galaxy, QSO or star) and R refers to a
random point (mock galaxy or QSO).  $DD$, then, could be the total number of
galaxy-QSO, QSO-galaxy, galaxy-star or star-galaxy pairs, to name a few
constructions.  If $DD$ were a reference to the number of QSO-galaxy pairs, 
then $DR$ would be the number of QSO-random pairs, where the random
catalogue consisted of a mock galaxy distribution.  The parameters $\bar{n}_D$ and
$\bar{n}_R$ refer to the total number density of data points and the total
number density of points derived from a random catalogue, respectively.
The term $\theta$ refers to the angular separation between the pairs.

Random catalogues are constructed by randomly creating points with
the same overall distribution as the sources.  For example, a random galaxy
catalogue contains points randomly distributed across the sky except
where there are holes in the APM or SDSS and a
random QSO catalogue contains points distributed according to the
coverage map of the 2QZ. The 2QZ coverage map is produced by
calculating the ratio of spectroscopically observed objects to input
catalogue objects in each region defined by the intersection of 2dF
pointings.  This distribution is then pixelized using $\rm{arcmin}^2$ pixels 
\cite{Cro01}.  When estimating the correlation
function, the random catalogue is always at least
50 times bigger than the data catalogue.
Note that the cross-correlation between, say, galaxies and QSOs, could
be performed in one of two directions, by centring on QSOs and counting
galaxies or by centring on galaxies and counting QSOs.  With ideal
samples, these procedures should be equivalent.  In many
cases, if gradients or biases that are not accounted for in the
random catalogue exist in one or both samples, the two directions
may not be exactly equivalent.

Since we are mainly interested in small scale ($\sim 10$~arcmin)
cross-correlations, we measure them locally by splitting both the NGC
and SGC into 15 individual fields.  In the north, this is purely arbitrary but 
in the south, the field boundaries correspond to the edges of the APM
plates.  Correlations are then counted within each
individual field, with the resulting numbers of pairs being totalled
for all fields to yield the total number of pairs across the entire survey
area. The total number of data-data (DD) pairs and data-random (DR) pairs are 
then taken in ratio, as per Equation (5), to estimate the global
correlation function.   This field-to-field
analysis should nullify the effects of different photometric zero-point
calibrations between plates, or gross variations across strips.  Errors
are determined via the spread in
the local correlation function between each field.  Essentially, these
field-to-field errors are $1\sigma$ standard deviations in the value of
the correlation function between fields, inverse variance-weighted to
account for the different numbers of sources on each field.  Errors are
thus estimated via

\begin{figure} 
\begin{centering}
\begin{tabular}{c}
{\epsfxsize=8truecm \epsfysize=8truecm \epsfbox[55 170 540 600]{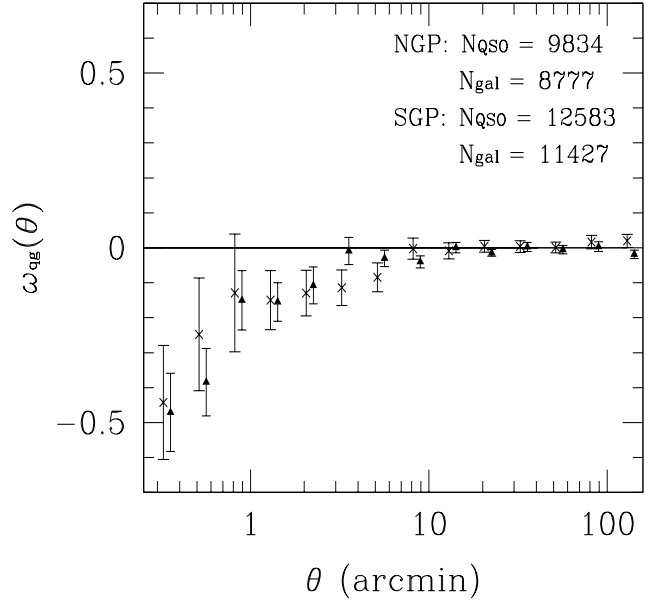}}
\end{tabular}
\caption
{The cross-correlation between 2QZ QSOs, and galaxies of limiting magnitude $B=20.5$
  found in objectively derived groups of at least seven members, for
  both 2QZ strips.  Crosses correspond to the NGC strip, triangles to
  the SGC strip.  The numbers of each sample within the boundaries of
  the 2QZ are displayed.  Errors are field-to-field.}
\label{fig:galcorrNGP.ps}
\end{centering}
\end{figure}

\begin{figure} 
\begin{centering}
\begin{tabular}{c}
{\epsfxsize=8truecm \epsfysize=8truecm \epsfbox[55 170 540 600]{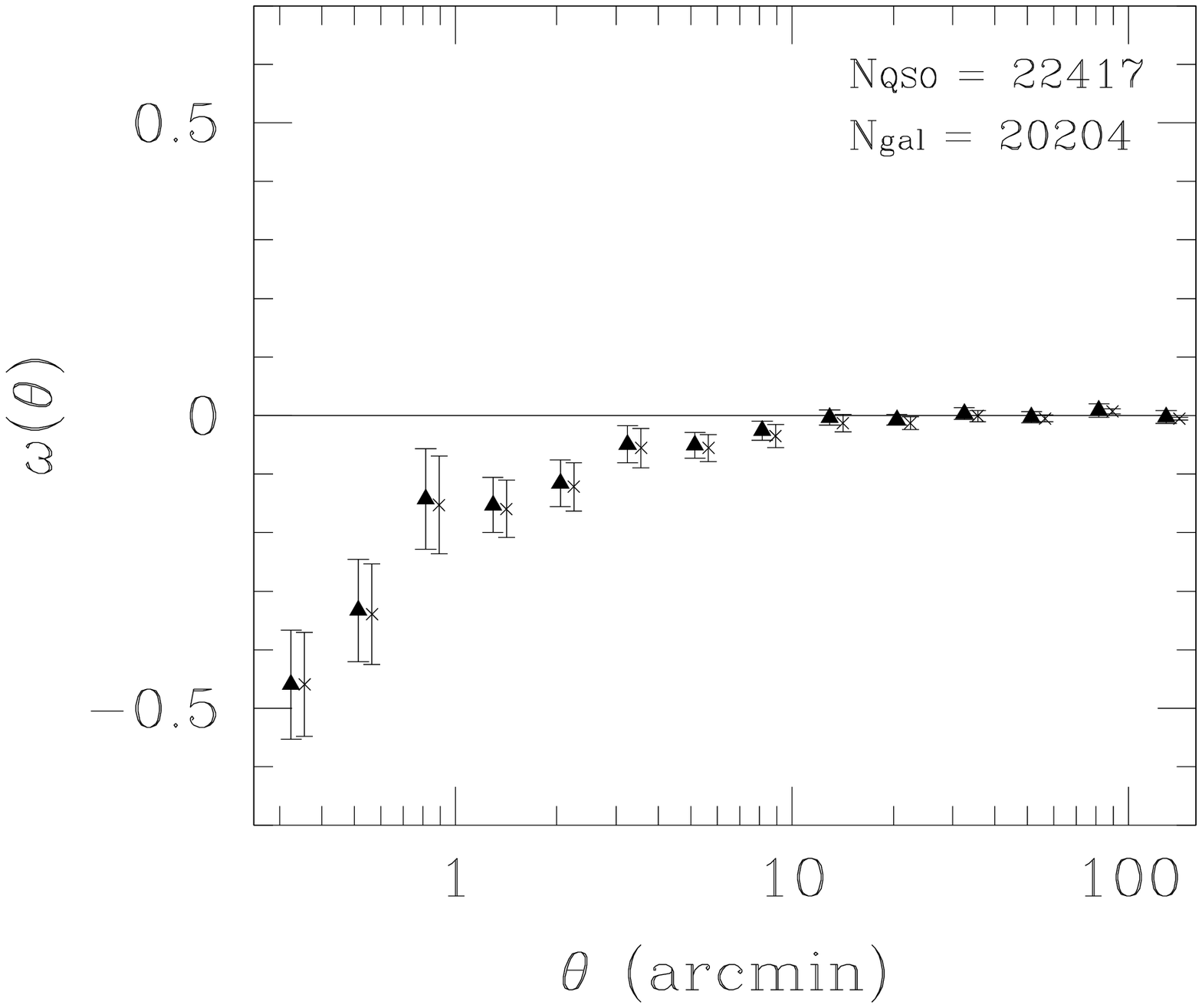}}
\end{tabular}
\caption
{The cross-correlation between galaxies of limiting magnitude $B=20.5$
  found in objectively derived groups of at least seven members and 2QZ QSOs,
  combined for both the southern and northern 2QZ strips.  Both results
  centring on QSOs and counting galaxies (triangles) and centring on
  galaxies and counting QSOs (crosses) are presented.  The numbers of each sample
  within the boundaries of the 2QZ are displayed.  Errors are field-to-field.}
\label{fig:galrevcomb.ps}
\end{centering}
\end{figure}

\begin{figure} 
\begin{centering}
\begin{tabular}{c}
{\epsfxsize=8truecm \epsfysize=8truecm \epsfbox[55 170 540 600]{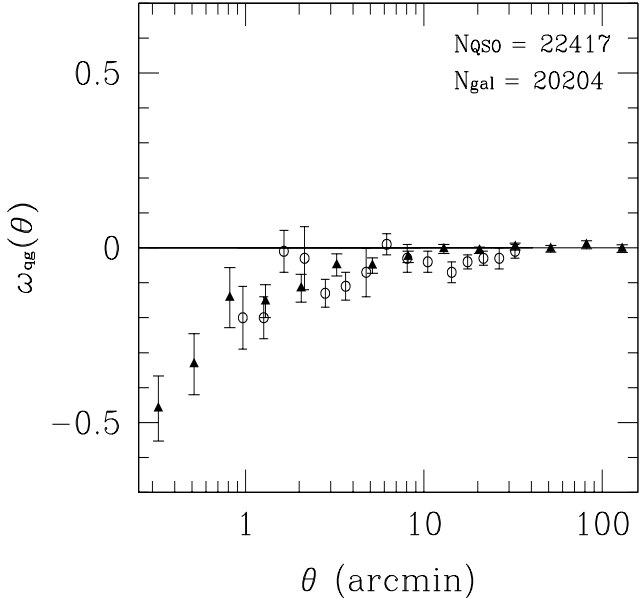}}
\end{tabular}
\caption
{The comparison between Figure 7 and the result of BFS88.  The BFS88
  data (open circles) have been scaled to account for their stellar contamination.
}
\label{fig:galrevcomb.ps}
\end{centering}
\end{figure}

\begin{eqnarray}
\sigma_{\omega}^2 =
\frac{1}{N-1}\sum_{L=1}^{N}\frac{DR_L(\theta)}{DR(\theta)}(\omega_L(\theta) 
- \omega(\theta))^2
\end{eqnarray}

where $\sigma_{\omega}$ is the field-to-field error on the correlation
function, and the subscript $L$ stands for local, referring to an
individual plate.  The $DR_L/DR$ factor, then, weights each field so
that fields with more objects are lent more significance in the error calculation.

We have compared this field-to-field error estimate to Poisson error,
as estimated by $(1+\omega(\theta))/\sqrt{DD}$.  Although the two types of error estimate show a strong
relation, the Poisson estimate overestimates the error, as compared
to the variance in random simulations of the data.  This might be expected as an individual
source may be associated with structure at several separations, so errors
over a given range can be strongly correlated.  On the smallest scales,
especially on scales where there are no data-data pairs on
\textit{some} of the plates, even the field-to-field estimate of error 
breaks down.  When fitting models to the data, we use an
estimate of the error based on the variance between many random
simulations of the data weighted to reproduce the model of interest.

In the following subsections, we calculate a variety of field-to-field cross-correlation
functions using Equation (5) for 2QZ QSOs against groups of APM or
SDSS galaxies.

\subsection{Cross-correlation of QSOs and group galaxies}

In Figures 6 (NGC and SGC) and 7 (combined) we display the
cross-correlations between spectroscopically identifed 2QZ QSOs and
galaxies in groups of at least 7 members objectively derived from the
SDSS EDR in the NGC strip and the APM catalogue, in the SGC strip.  In Figure
7, we show both directions of correlation, we have centred on
galaxies in groups and counted QSOs,  and have centred on QSOs and
counted galaxies.  Figure 8 shows the comparison between our result and 
BFS88, where the BFS88 data have been scaled to allow for their
projected 25~per~cent contamination by stars. The results are displayed at the
smallest scales for which we still believe the field-to-field-errors.
The largest scale displayed is a few
bins before edge effects caused by the 2QZ strips only being
300~arcmin in declination begin to have any effect.  The numbers
displayed on each plot are the numbers of objects from each sample
present within the confines of the 2QZ boundary.  Note that the SDSS
EDR sample is half the size of the northern 2QZ strip, so does not
contribute as significantly to the combined signal as the APM.  A
redshift cut has been made at $z = 0.4$ in the QSO sample; if the
signal is caused by lensing, this will theoretically reduce the overlap 
between QSOs and foreground matter to at most 0.4~per~cent (see Section
2, above).

\begin{figure} 
\begin{centering}
\begin{tabular}{c}
{\epsfxsize=8truecm \epsfysize=8truecm \epsfbox[55 170 540 600]{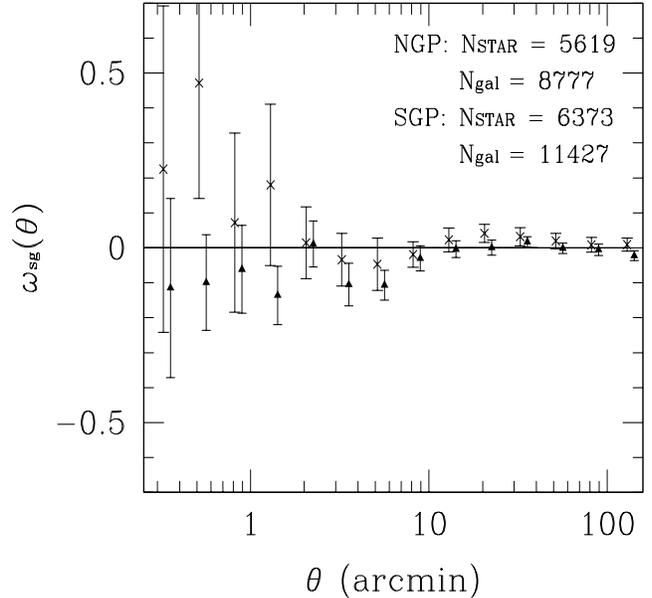}}
\end{tabular}
\caption
{The cross-correlation between 2QZ stars, and galaxies of limiting magnitude $B=20.5$
  found in objectively derived groups of at least seven members, for
  both 2QZ strips.  Crosses correspond to the NGC strip, triangles to
  the SGC strip.  The numbers of each sample within the boundaries of
  the 2QZ are displayed.  Errors are field-to-field.}
\label{fig:galcorrNGP.ps}
\end{centering}
\end{figure}

\begin{figure} 
\begin{centering}
\begin{tabular}{c}
{\epsfxsize=8truecm \epsfysize=8truecm \epsfbox[55 170 540 600]{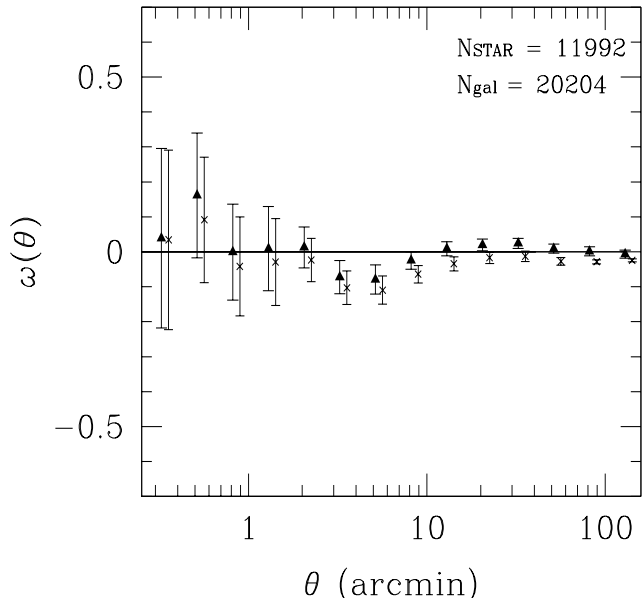}}
\end{tabular}
\caption
{The cross-correlation between galaxies of limiting magnitude $B=20.5$
  found in objectively derived groups of at least seven members and 2QZ stars,
  combined for both the southern and northern 2QZ strips.  Both results
  centring on stars and counting galaxies (triangles) and centring on
  galaxies and counting stars (crosses) are presented.  The numbers of each sample
  within the boundaries of the 2QZ are displayed.  Errors are field-to-field.}
\label{fig:galrevcomb.ps}
\end{centering}
\end{figure}

Figure 6 shows that there is good consistency between the northern and
southern correlation functions for QSOs versus galaxies in groups, justifying combining the signals.
Figure 7 demonstrates that there is excellent consistency in the
cross-correlation signal between QSOs and group galaxies irrespective
of the direction in which the function is calculated, suggesting
the signal is robust, free from the influence of any gradient or
incompleteness in the samples used.  There is a significant $3.0\sigma$
anti-correlation between galaxy groups and spectroscopically identified 
2QZ QSOs on scales out to 10~arcmin, based on collecting the data in a
single 10~arcmin bin and calculating the rms field-to-field variation.
For an Einstein-de-Sitter cosmology, 10~arcmin would translate to $1~
h^{-1}~\rm{Mpc}$ at an average group redshift of $z \sim 0.15$ ($1.1~
h^{-1}~\rm{Mpc}$ for $\Lambda CDM$.
The average anti-correlation in such a 10~arcmin bin is $-0.049$.  Our
data compare well in Figure 8 with the anti-correlation discovered by
BFS88 when correlating a UVX object sample with cluster galaxies.
BFS88 declared a more significant $4\sigma$ signal on $<$4~arcmin scales.
We have also measured the same effect after correcting the random QSO
catalogue for Galactic dust, using Schlegel maps \cite{Sch98}.  Although the
\textit{strength} of the effect is not altered, the variation between fields is
reduced, meaning the \textit{significance} of the result increases
slightly to $3.3\sigma$.  Note that, as neither of the galaxy
populations we use are corrected for Galactic dust, no correction for Galactic 
dust in the QSO sample should be necessary when cross-correlating the
2QZ with the galaxy populations.  In any case, correcting for Galactic
dust has a very small effect.

\begin{figure*} 
\begin{centering}
\begin{tabular}{c}
\centerline{\epsfig{figure=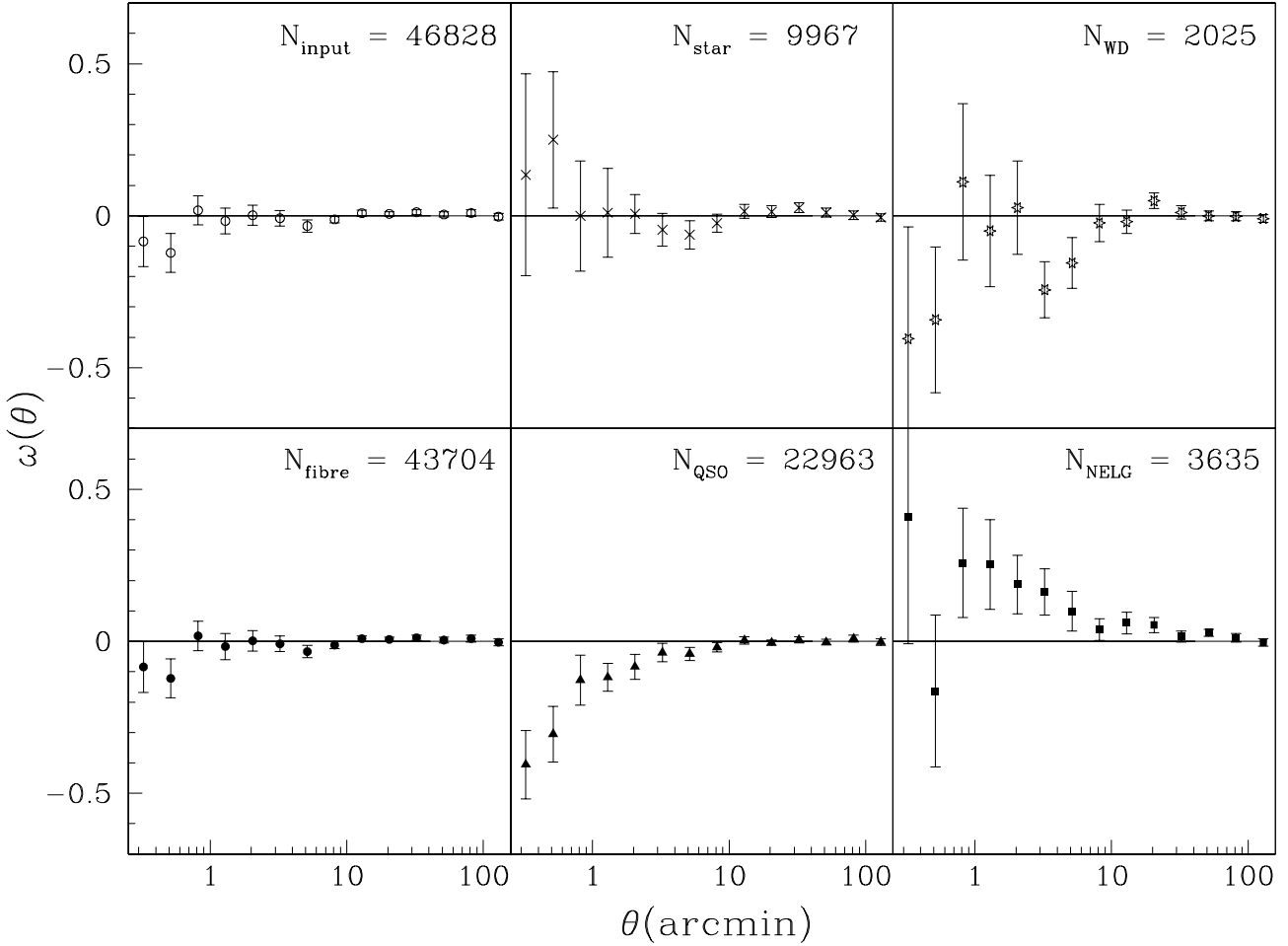,width=5.2in,height=7in,angle=-90}}
\end{tabular}
\caption
{The cross-correlation between 20204 galaxies of limiting magnitude $B=20.5$
  found in objectively derived groups of at least seven members and
  samples of objects taken from the 2QZ, combined for both the northern
  and southern 2QZ strips.  The upper-left panel is the input
  catalogue.  The lower-left panel is the fibre catalogue.  The other
  panels are specific objects that comprise the majority of the fibre catalogue; stars, 
  QSOs, White Dwarfs and NELGs.  The numbers of each subsample are
  displayed.}
\label{fig:components.ps}
\end{centering}
\end{figure*}

To test the anti-correlation between group galaxies and QSOs, we need a 
control sample that has been through the same measurement and reduction
as the QSO sample.  BFS88 found no correlation between a large ($\sim$ 27,000) 
control sample of non-UVX stars and group galaxies, however, our situation is notably more
complicated.  As we require a control sample that has been through the same
processes as the QSOs, including spectroscopic identification, we are
restricted to a smaller sample of stars ($\sim$ 12,000) compared to 
our QSO sample ($\sim$ 22,500). Additionally, the selection criterion
of the 2QZ tends to pick out specific populations of stars, such as White Dwarfs.

In Figures 9 (NGC and SGC) and 10 (combined) we show the
cross-correlation between spectroscopically identified 2QZ stars and
galaxies in groups.  Unlike the QSO result, the NGC and SGC
cross-correlations are not entirely consistent, suggesting that the
different physical distribution of stars in the two strips effects
the correlation function.  The combined results are also inconsistent,
although the errors are sizable.  The combined results show evidence of 
gradients on large scales, as, unlike in the case of the QSOs,  the
cross-correlation result is dependent on whether we centre on stars and
count galaxies or centre on galaxies and count stars.  Indeed, the distribution of
stars within our galaxy does display a gradient with galactic latitude
in both the NGC and SGC, unlike the distribution of QSOs, which is
flat.  The cross-correlation between stars and galaxies has some
negative points on 3-7~arcmin scales but does not have the same form as 
the QSO anti-correlation.  In particular, the star-signal does not
continue to decrease on scales less than 3~arcmin.  If we calculate the 
significance of the signal out to 10~arcmin, we find a $1.6\sigma$
($1.4\sigma$ with correction for dust) anti-correlation for the result
centring on stars and counting galaxies and a $3.0\sigma$
($2.9\sigma$ with dust-correction) anti-correlation for the result
centring on galaxies and counting stars.  Much of this discrepancy is
caused by the large-scale gradients; if the galaxy-centred result is
moved up so there is no anti-correlation on large scales, it comes into 
line with the star-centred result.  The anti-correlation in both of the 
combined star results is caused entirely by the few points on 3-7~arcmin
scales.  Surprisingly, most of the anti-correlation between stars
and group galaxies is actually caused by White Dwarfs.  If we discard
the White Dwarfs from the stellar sample, there is no significant
anti-correlation between stars and group galaxies (see Figure 11).
Whether there is a physical reason for this (circumstellar dust?)
remains to be seen, otherwise the most likely explanation for the
anti-correlation seen at 3-7~arcmin in both the stars and White Dwarfs
is a statistical fluctuation due to the low numbers of these objects.

We regard the results using the above control samples of stars as
encouraging in terms of ruling out a systematic source for the QSO-group
anti-correlation. However, because of the low numbers of stars, their
gradients in Right Ascension and the anti-correlation detected at 3-7~arcmin, stars may
not form the ideal control sample and there may be a residual doubt as to
whether there is a systematic contribution to the QSO
anti-correlation on 3-7~arcmin scales caused by the fibre positioning
constraint of the 2dF instrument.  After all, 2dF observed the $b<19.5$ galaxies and QSOs
simultaneously and in dense fields close pairs of objects may have been
missed due to the minimum fibre spacing, even though 2dF candidates were
given a higher priority in the fibre allocation to prevent imprinting
the galaxy structure on the QSO distribution. It is easy to show there
is no fibre positioning effect by comparing the cross-correlation of 
the 46,000 objects in the 2dF input catalogue with the 43,000 objects that 
were observed spectroscopically. These results are shown respectively in 
the upper left-hand and lower left-hand panels of Figure 11.  These 
two results are in all respects identical with no systematic difference 
between them, leading us to conclude that there is no anti-correlation
induced on these scales by the fibre positioning constraint. Therefore we 
conclude that the QSO-group galaxy anti-correlation is probably real.

Note that the reason the correlation between group galaxies and the input
catalogue is flat, even though it contains the significantly anti-correlated QSO
signal, becomes apparent when we split the input catalogue up into its
constituent parts.  The right-hand
four panels in Figure 11 display the main subsamples of the input
catalogue.  We can see that the Narrow Emission Line Galaxies included in the input
catalogue exhibit a positive correlation that cancels out the
anti-correlation exhibited by QSOs, leaving the input catalogue result flat.

\subsection{The dust hypothesis}

BFS88 originally attributed the anti-correlation between QSOs and
galaxies in groups to interim dust in clusters, finding that an
absorption in the $b-$band of $A_b \sim$ 0.2~mag was sufficient to cause
their observed anti-correlation.  Such absorption is at the upper limit 
allowed by Ferguson (1993), who quoted a maximum reddening of $E(B-V) 
\leq 0.06$  from a composite study of the Mg$_2$ index of 19 nearby
clusters and rich groups.  Using colour information provided by the 2dF survey, we can
limit the culpability of dust in causing the anti-correlation signal.
The 2QZ measures $u-b_j$ and $b_j-r$ colours.  If the anti-correlation between
QSOs and group galaxies were due to dust, we would expect to observe a
complementary reddening of QSOs.

\begin{figure} 
\begin{centering}
\begin{tabular}{c}
{\epsfxsize=8truecm \epsfysize=8truecm \epsfbox[55 170 540 600]{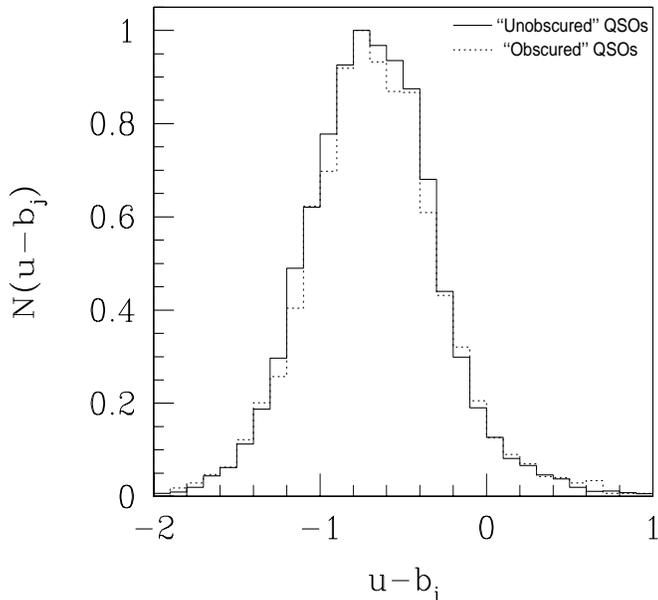}}
\end{tabular}
\caption
{The relative distribution of $u-b_j$ colours of 2QZ QSOs.  The solid histogram shows the 
  colours of QSOs that \textit{do not} lie within 10~arcmin of a
  group centre.  The dashed histogram shows the colours of QSOs that
  lie within 10~arcmin of a group centre.  
}
\label{fig:colourub.ps}
\end{centering}
\end{figure}

\begin{figure} 
\begin{centering}
\begin{tabular}{c}
{\epsfxsize=8truecm \epsfysize=8truecm \epsfbox[55 170 540 600]{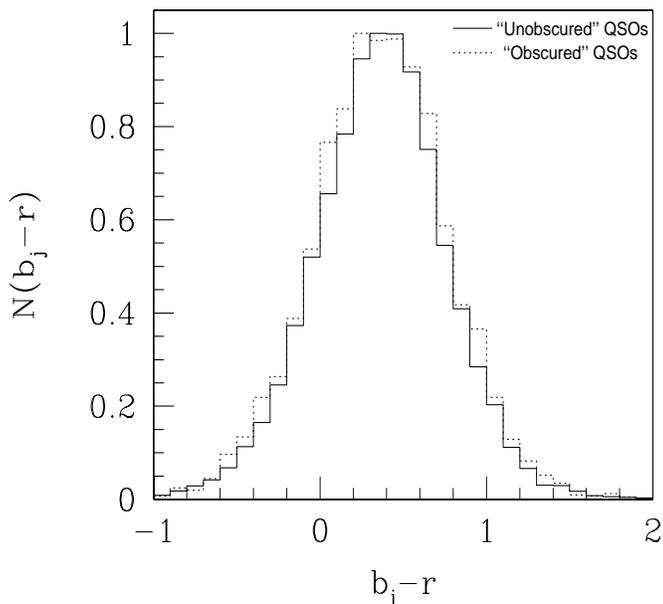}}
\end{tabular}
\caption
{The relative distribution of $b_j-r$ colours of 2QZ QSOs.  The solid histogram shows the 
  colours of QSOs that \textit{do not} lie within 10~arcmin of a
  group galaxy.  The dashed histogram shows the colours of QSOs that
  lie within 10~arcmin of a group galaxy.  
}
\label{fig:colourbr.ps}
\end{centering}
\end{figure}

In Figures 12 and 13 we show the distributions of colours of 2QZ QSOs that lie within
10~arcmin of any group centre (`obscured' QSOs)  and QSOs that do not
lie within 10~arcmin of any group centre (`unobscured' QSOs).  Figure
12 is for $u-b_j$ colours, Figure 13 for $b_j -r$ colours.  We have
selected 10~arcmin as the radius of interest as it corresponds to the
extent of the anti-correlation signal in Figure 7.  Were the
anti-correlation due to intervening dust in galaxy groups, as proposed
by BFS88, we would expect to see a complementary reddening of QSOs on
the scale of the anti-correlation, and the distributions of `obscured'
and `unobscured' QSOs would differ.

A two-sample Mann-Whitney U-test fails to reject the null hypothesis that the
`obscured' and `unobscured' distributions have the same median, for
both the $(b_j-r)$ and $(u-b_j)$ colours.  A two-sample Kolmogorov-Smirnov test fails to reject 
the null hypothesis that the `obscured' and `unobscured' distributions
are drawn from the same parent population, again for both colours.
Student's t-test demonstrates that the means of the `obscured' and
`unobscured' samples are in agreement.  For the $(u-b_j)$ colour
distributions, the `unobscured' mean and standard error are $-0.6789
\pm 0.0030$ and for the `obscured' are $-0.6687 \pm 0.0064$.  For the
$(b_j-r)$ colour distributions, the `unobscured ' mean and standard
error are $0.3644 \pm 0.0030$ and the `obscured' mean and standard
error are $0.3626 \pm 0.0065$.  There are 4025 `obscured' QSOs and
17752 `unobscured QSOs.  Student's t-distribution sets the following 95 (99)
per~cent upper limits on reddening between the two distributions:
$E(u-b_j) = 0.012$ ($E(u-b_j) = 0.016$), $E(b_j-r) = 0.012$ ($E(b_j-r)
= 0.016$).  Now, our average group size is around 2.5~arcmin and larger
groups in our sample have an angular size of 5~arcmin, so we
might also be interested in any reddening on these scales.  Repeating
the above analysis on these scales, Student's t-test suggests the 95 (99) per~cent reddening limits
between the QSO population within 5~arcmin of any group centre and the
QSO population lying greater than 10~arcmin are $E(b_j-r)=0.020$ $(0.028)$ and
the 95 (99) per~cent reddening limits between the QSO population within
2.5~arcmin of any group centre and the remainder are $E(b_j-r)=0.039$
$(0.056)$.  These reddening limits apply for both the $u-b_j$ and
$b_j-r$ colours.  The limits inevitably increase as the `obscured' population 
dwindles in size, although there are still 286 QSOs within 2.5~arcmin
of any group centre.  Again, for both the 2.5~arcmin and 5~arcmin
scales, the Kolmogorov-Smirnov test failed to reject the hypothesis
that the `obscured' and `unobscured' distributions are drawn from the
same parent distribution of colours.  Assuming the usual Galactic dust
law, the 95~per~cent limit from the $b_j-r$ colours on B-band absorption within 5~arcmin
($\sim~0.5~h^{-1}~\rm{Mpc}$ at $z \sim 0.15$) of any group
centre is $A_B < 0.06$~mag.  The similar limit within 2.5~arcmin of any
group centre is $A_B < 0.13$~mag.
A simple model (see BFS88 for details), taking the slope in the QSO
number magnitude counts to be 0.29, suggests that these levels of
absorption correspond to a correlation function of $\omega_{cq} <
-0.039$ for $A_B < 0.06$~mag and $\omega_{cq} < -0.077$ for $A_B <
0.13$~mag.  For the entire sample out to 10~arcmin, the 95~per~cent reddening
limits suggest an anti-correlation of only $-0.029$.

It seems that 2QZ QSOs close to galaxy groups are insufficiently reddened to explain the
anti-correlation signal as an effect of dust in galaxy groups.  It
might be argued that as the 2QZ is colour-selected, our reddening
values are biased by the limits set on the colours of 2QZ QSOs.  We do
not mean to argue that our values are objective determinations of the
extent of dust in galaxy groups, only that there is insufficient
reddening of QSOs close to galaxy groups within the 2QZ to explain the
anti-correlation signal.  Undoubtedly, there will be some heavily reddened QSOs
close to group centres that the 2QZ fails to observe, however, the QSOs 
the survey \textit{does} observe have no tendency to redness close to
group galaxies.  The 2QZ sample colour-colour distribution peaks significantly 
bluewards of any colour limit, so we believe that the low reddening 
measure is not forced by the colour selection procedure.  

Previous evidence for dust in galaxy groups has been controversial,
with some authors claiming detections and others claiming upper
limits.  Girardi et al. (1992) confirm galaxies in local groups that
are blueshifted relative to the group average tend to have a larger
colour excess than that group average, and suggest background
galaxies falling towards the group centre suffer reddening by dust in
the group.  The amount of dust Giardi et al. (1992) infer is $E(B-V)$
$\sim 0.1 - 0.2$ on 0.75~Mpc scales, equivalent to an absorption of $A_B \sim 0.4 -
0.8$~mag (for the usual galactic dust law).  Such an amount of absorption 
is at odds with the upper limit quoted by Ferguson (1993).  Ferguson
studied the correlation of Mg$_2$ index with $B-V$
colour for local groups and clusters, and compared these results to a
sample of field galaxies.  Ferguson (1993) quoted $E(B-V) \leq 0.06$ as
a conservative (90~per~cent) upper limit on reddening in a sample of clusters and
groups and, similarly, an upper limit of $E(B-V) \leq 0.05$ ($A_B <
0.2$~mag) for a sample of poorer groups.  Ferguson (1993) considers
a scale of 0.5~Mpc to be the central contribution to his reddening limits.  The 95~per~cent upper limits
on our  reddening results out to $0.5~h^{-1}~\rm{Mpc}$ (5~arcmin) allow for us much absorption
by dust as $A_B < 0.06$~mag, more consistent with the results of Ferguson
(1993) rather than those of Girardi et al. (1992).

The possibility remains that some specifically tailored dust model could still
explain the anti-correlation.  One possibility is a smooth distribution
of dust around galaxy groups that does not obey the reddening laws
observed in our galaxy (so-called `grey dust').  A less ad hoc
explanation is a patchy distribution of dust around galaxy
groups that is heavily concentrated in the line of sight direction, such that
QSOs are completely obscured without being reddened.  Averaged over
many groups, this would have the effect of removing QSOs from the
sample near galaxy groups without overtly reddening the QSO sample on
similar scales.  Finally, we must consider the possibility that
multiple effects of dust and lensing combine to produce the observed
anti-correlation.  So, although our current analysis can rule out a smooth
distribution of typical dust around galaxy groups, we cannot deny more
tailored dust models, such as grey dust, patchy dust or a combination
of gravitational lensing and dust.

\section{Weak lensing}

We have seen that the $3.0\sigma$ anti-correlation between 2QZ QSOs and
galaxies in groups is unlikely to be either a selection effect or a
product of normal, smoothly distributed dust in galaxy groups obscuring QSOs.  In this
section we consider the possibility that the anti-correlation results
from the statistical lensing of QSOs by foreground galaxy groups \cite{Cro99}.

\begin{figure} 
\begin{centering}
\begin{tabular}{c}
{\epsfxsize=8truecm \epsfysize=8truecm \epsfbox[55 170 540 600]{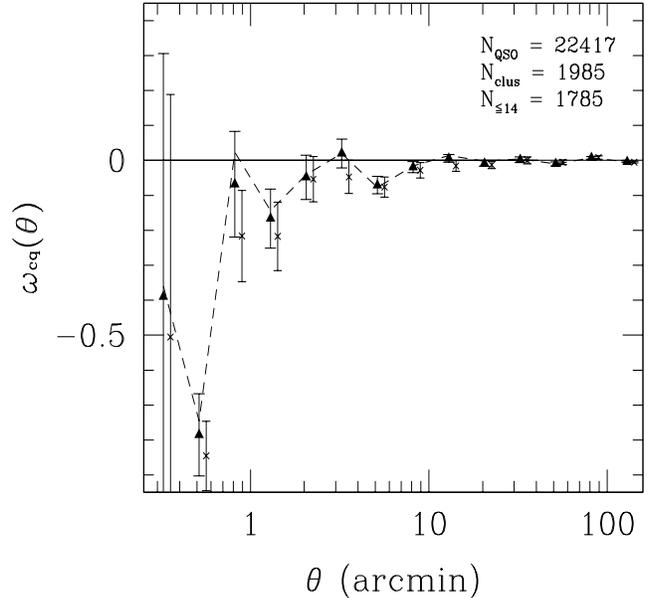}}
\end{tabular}
\caption
{The cross-correlation between 2QZ QSOs and objectively derived galaxy
  groups of limiting magnitude $B=20.5$ combined for both the southern 
  and northern 2QZ strips.  The numbers of each sample within the
  boundaries of the 2QZ are displayed.  Crosses are derived by
  weighting each galaxy group by the number of galaxies in the group
  and have been offset slightly for display purposes.  Triangles are
  derived counting each group equally.  Errors are field-to-field.
  Also marked as a dashed line is the unweighted result for only those groups with
  14 or fewer members.  Errors on this line are as plotted for the triangles. }
\label{fig:centres.ps}
\end{centering}
\end{figure}

\subsection{Modelling approach}

In the Appendix, we outline models we use to describe how the
statistical lensing of background QSOs may trigger an anti-correlation
signal of the type discovered above.  The models describe a slightly
different situation than hitherto discussed, the lensing of background
flux around the \textit{centres} of dark matter
profiles, and so describe the correlation function of QSOs against the
\textit{centres} of galaxy groups.  Previously, to compare our results
with BFS88, we have measured the correlation between QSOs and
individual galaxies in groups.  To facilitate comparison with models,
we now cross-correlate QSO positions with group centres, rather than
group galaxies.  Additionally, in prior Sections larger groups of
galaxies have been weighted more highly as each individual galaxy position within the
group would be counted.  We fit models to data where the correlation function is unweighted -
each cluster is considered equally.

The cross-correlation between 2QZ QSOs and the centres of galaxy groups
objectively determined from the APM South and the SDSS EDR is displayed 
in Figure 14.   For comparison, both the weighted and unweighted
cross-correlations between QSOs and the centres of galaxy groups are
shown.  The weighted result (crosses) is a reflection of the analysis
in Section 4.2, the QSOs are counted against each galaxy in a group, so 
larger groups are lent more significance.  The unweighted result is not 
biased by the size of the groups.  Binning the
data in a sole bin out to 10~arcmin and measuring the rms
field-to-field variation in this bin, the average anti-correlation of
the \textit{weighted} result is $-0.049$ with a significance of $2.9\sigma$.
Comparing with the results in Section 4.2 , the strength and significance of the result
weighted by galaxy number proves essentially identical whether we
correlate QSOs with galaxies in groups or the
group centres.   The \textit{unweighted} result has an average
anti-correlation out to 10~arcmins of $-0.034$ with a significance of $2.9\sigma$.
When we do not weight the cross correlation by the number of galaxies
in the group, the \textit{strength} of the result thus drops by
30~per~cent compared to the anti-correlation between QSOs and
groups outlined in Section 4.2.  We can deduce that the
anti-correlation signal is stronger for larger groups, as would be
expected if it is due to lensing.  In fact, it is straightforward to
calculate that if we assume a linear relationship between the mass of a
group and the number of galaxies in a group we would expect just such a
30~per~cent drop.   To further demonstrate that the cross-correlation
signal is not biased up by the handful of large clusters in our sample,
we have cross-correlated QSOs against a sample of galaxy groups that
has the largest 10~per~cent of objects removed.  The resulting sample
of galaxy groups has between 7 and 14 galaxies per group.  The
anti-correlation for this sample is plotted as the dashed line in
Figure 14 and has essentially identical error bars to the full
(unweighted) sample (marked with triangles).  There is practically no
difference in the result if we remove the largest 10~per~cent of
objects from our group sample, the anti-correlation signal thus derived 
has a strength of $-0.032$ and a significance of $2.5\sigma$.

We model the lensing groups as dark matter haloes, either Singular
Isothermal Spheres or NFW profiles (as described in the Appendix).  In the case of the SIS, the free
parameter is the velocity dispersion of the sphere.  The free parameter 
in the NFW model is the mass within $1.5\mpc$ of the centre of the
halo centre.   For the lensing analysis, we use a faint-end QSO
number-count slope of $\beta = 0.29$
(see Section 2).  The overwhelming majority of QSOs
in the 2QZ lie fainter than the knee of a broken power-law model (only 10~per~cent
are brighter than $b_j = 19.1$.  We have reproduced models both
approximating the number-magnitude
counts as a single power-law with slope $\beta = 0.29$ and using the
full smoothed power law determined in Section 2 (see Equation 1) and
find no significant difference between the two approaches.  

The lens and source distances required by the lensing models (see
Equation A6, for example) are calculated by randomly sampling
redshifts from the distributions displayed in Figure 3 and deducing the 
average separation of a galaxy-QSO pair, as outlined in Section 2.  
The redshifts are translated to distances using an Einstein-de Sitter cosmology.

When fitting models, we use an estimate of the error based on 500
`mock' QSO catalogues of the same size and completeness
as the 2QZ, calculating the cross-correlation between these
mock catalogues and our galaxy group centres in the normal way, and
then determining the rms errors between the catalogues.   We fit a
reasonable model to the anti-correlation result, then distribute the
random placement of QSOs in the
mock catalogues to reflect that model.  So, if the cross-correlation fit
has a value of -0.1 at 2~arcmin, a random QSO that is generated
2~arcmin from a group centre is only 90~per~cent as likely to be included
in the mock QSO catalogue as one a long way from any group.
Similarly, a QSO that then lies within 2~arcmin of 2 group centres is
81~per~cent as likely to be included. We have tested the independence of 
these errors by measuring the covariance of points in adjacent bins
averaged over the 500 simulations, finding the covariance
insignificant on all scales.

\subsection{Model fits}

Figure 15 shows the best-fit models for the SIS and NFW lensing haloes
obtained by minimising the $\chi^{2}$ statistic.
The errors are calculated from the standard deviation in the
anti-correlation signal of 500 mock QSO catalogues as outlined
above, with `mock' QSOs distributed according to a reasonable
model.  In the SIS case, mock QSOs are distributed according to a
$\sigma = 600~\rm{km~s^{-1}}$ model when calculating errors.  In the
NFW case, mock QSOs are distributed according to a $M_{1.5} = 3 \times
10^{14}~h^{-1}~\msun$ model when calculating errors.  For brevity, the
models are both displayed in Figure 15 against the error bars
calculated for the NFW case.  We
determine the best-fitting models out to 10~arcmin, the extent of the
anti-correlation.  Once the best-fitting model is determined, the
errors are scaled so the reduced $\chi^2$ is equal to 1 and then 1$\sigma$ error-bars
on the best-fitting model are calculated from this renormalised $\chi^2$ distribution.
The best-fit SIS has a velocity dispersion of $\sigma = 1156 \pm ^{93}_{327}~\rm{km~s^{-1}}$ with a
reduced $\chi^2$ of 0.8.  The best-fit NFW has a mass of
$M_{1.5} = 1.2 \pm{0.9}  \times 10^{15}~h^{-1}~\msun$ with a
reduced $\chi^2$ of 1.2.  The data cannot distinguish variations in the
NFW $\gamma$ parameter in the range $\frac{1}{3}<\gamma<1$ (see
Equation A11).  As $\gamma$ increases, the predicted anti-correlation
decreases below 1~arcmin and a test between these cases should be 
possible in bigger datasets.  In the current datasets there is hardly 
enough power to distinguish between the best-fitting SIS and NFW models.

The Einstein Radius of the best-fitting SIS model is around 30~arcsec,
which results in a radical dip in the solid line in Figure 15, corresponding to 
the terms in the denominator of Equation A6 being equal.  At
separations within the Einstein Radius, we enter the strong lensing
regime and the SIS model predicts that each source QSO will produce two 
images.  One of these images has already been covered in the SIS model
prediction for scales larger than the Einstein Radius.  The second
image appears within the Einstein Radius and might be considered a further 
prediction of the SIS model.  The divergent nature of the SIS density
profile on small scales means the dip in Figure 15 may be predicted to
occur at a scale which is unphysically large.  Details of the model 
near the Einstein Radius do not affect the fitted SIS masses much,  as
evidenced by their similarity to those masses determined from NFW
fitting where the Einstein Radius is much smaller.  The full
consequences of strong lensing in the 2QZ, including an analysis of the 
numbers of multiply-lensed QSO systems, are discussed elsewhere (by Miller 
et al. 2003).

As most of our model analysis is made in the weak lensing regime, we might be
wary of any fit to the smallest scale data points in the SIS case.   As
a consistency check, we have used mock QSO
catalogues to make a direct test of the significance of rejection of 
$\sigma = 600~\rm{km~s^{-1}}$ and $\sigma = 1140~\rm{km~s^{-1}}$ SIS models, since the
mock catalogues were produced for these specific cases.  We determine
how often cross-correlating mock QSO catalogues with galaxy groups can return an
anti-correlation of significance $-2.9\sigma$, as found for the real 2QZ data.  We display
this result in Figure 16.  We have created 250 mock QSO catalogues with the same
size and completeness as the 2QZ.  These are then cross-correlated
against our galaxy groups and the strength of the cross-correlation 
is measured for each mock QSO catalogue.  The measure of the significance
of a cross-correlation is as we have used throughout this paper, based
on counting the data in one bin out to 10~arcmin.  The mock QSO catalogues may
also be distributed, as outlined above, to reflect various models for the
lensing halo.  We find that an anti-correlation of $-2.9\sigma$ is
measured 0/250 times if there are no lensing haloes, only 7/250 times ($2.2\sigma$) if the
lensing haloes are modelled as an SIS with a $600~\rm{km~s^{-1}}$ velocity
dispersion (roughly equivalent to $\Omega_m = 0.3$) and 137/250 times for an SIS
with an $1140~\rm{km~s^{-1}}$ velocity dispersion (roughly equivalent to
$\Omega_m = 1$).   If we had chosen to plot the \textit{strength} of
the anti-correlation rather than the significance, we would find that a
$600~\rm{km~s^{-1}}$ model produces an anti-correlation of strength
-0.034 (as found for the data) only 13/250 times ($1.9\sigma$).  Either way, these
results reject the $600~\rm{km~s^{-1}}$ model at about the
5~per~cent significance level.   Note that this is a slightly stronger
rejection than the error bars quoted in the first paragraph of this
Section, which correspond to a $1.7\sigma$ rejection.  The discrepancy
is likely because renormalising the reduced $\chi^2$ statistic to 1 is not an
entirely fair representation of the error.  If this renormalisation is
not carried out, the rejection of the $600~\rm{km~s^{-1}}$ model rises
to $1.9\sigma$.

Wu et al. (1996) have pointed out that the inclusion of a uniform plane
in modelling a dark matter halo may be considered a good reflection of the lensing
influence of large-scale structure.  Following the model of Croom \&
Shanks (1999), we have considered the effects of including such a plane 
in our SIS model.  Including a plane of dark matter with our SIS has no 
influence on the best-fit model or it's error-bars.  In fact, the only
effect worth remarking is to slightly lower our rejection of low
velocity dispersion models.  The best-fit SIS model rejects
$600~\rm{km~s^{-1}}$ at a $1.7\sigma$ level.  When a plane is included,
this rejection drops to $1.5\sigma$.  If we consider the group centre
auto-correlation function ($\omega_{cc}$) of Stevenson, Fong and Shanks
(1988), then we might expect the clustering of groups to have little
effect on our signal.  Integrating under $\omega_{cc}$ out to 10~arcmin 
suggests only a 25~per~cent contribution to our signal from the
clustering of groups.

There is still some debate over a number of the parameters used in the
modelling process.  Changing the QSO number-count faint-end slope to
$\beta = 0.34$ raises the SIS model estimate of the
velocity dispersion by $\sim$ 10~per~cent and lowering the index to $\beta =
0.24$ lowers the estimate about 10~per~cent.  Changing to a Concordance
cosmology with $\Omega_m = 0.3$ and $\Lambda = 0.7$, to model
lens-source separations  raises the estimate by $\sim$ 5~per~cent.  In
the case of the NFW profile, lowering $\beta$ to
$0.24$ lowers the mass estimate by $\sim$ 30~per~cent, raising $\beta$ to $0.34$
raises the mass estimate by $\sim$ 50~per~cent and changing to a Concordance
cosmology raises the mass estimate by $\sim$ 5~per~cent.  In short, cosmology does
not really affect our estimates but the exact gradient of the faint-end 
slope of the QSO number-magnitude counts may well be important,
especially for the NFW profile.  Reducing the QSO number-count faint-end slope to
$\beta = 0.15$, a strong minimum value (being the
differential counts slope determined in Section 2.1) would reduce the
NFW mass estimate by 70~per~cent, bringing the NFW model mass
prediction in line with a $\Lambda$CDM cosmology.  It is possible that
the magnification values generated by groups
are large enough to reach so faintly into the QSO counts that the
value of $\beta$ is as low as the differential counts value.  This could
conceivably allow a $\Lambda$CDM cosmology combined with an NFW profile
to explain the anti-correlation between QSOs and galaxy groups.

\begin{figure} 
\begin{centering}
\begin{tabular}{c}
{\epsfxsize=8truecm \epsfysize=8truecm \epsfbox[55 170 540 600]{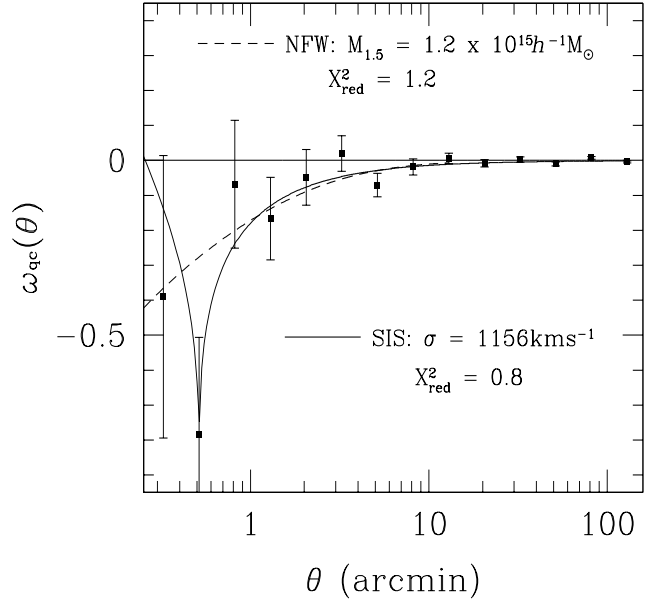}}
\end{tabular}
\caption
{The cross-correlation between 2QZ QSOs and objectively derived galaxy
  groups of limiting magnitude $B=20.5$ combined for both the southern 
  and northern 2QZ strips together with best-fit models.  Errors are
  standard deviations on the best-fitting
  NFW model derived from 500 mock QSO samples.}
\label{fig:unweight.ps}
\end{centering}
\end{figure}

\subsection{Estimating $\Omega_m$ in Groups}

In the Appendix, we also note how estimates of the average group mass, or 
velocity dispersion, can be turned into estimates of the mass density of
the Universe that is associated with groups (see Croom \& Shanks 1999).
The sky density of our groups is around
$0.85\deg^{-2}$ in both the NGC and SGC strips.  Note that
this is significantly larger than the density of Abell
clusters ($\sim 0.1\deg^{-2}$).  Croom \& Shanks (1999) have estimated
the typical space density of such groups as  $3 \pm 1
\times10^{-4}h^3\rm{Mpc^{-3}}$.  Using this value, Equation (A22) may be written

\begin{eqnarray}
  \Omega_m = (\frac{\sigma}{1125 \rm km s^{-1}})^2
\end{eqnarray}

assuming that the extent of the anti-correlation signal is $\theta = 10
\pm 2$~arcmin ($r=1 \pm 0.2 \mpc$).  Equation (A23) is equivalently

\begin{eqnarray}
  \Omega_m = \frac{M_{ \rm{NFW}}}{9.2 \times 10^{14}\it{h}\rm^{-1} \msun} \geq
  \frac{M_{1.5}}{9.2 \times 10^{14}\it{h}\rm^{-1}\msun}
\end{eqnarray}

Using our best-fit estimates for the SIS (including scales between 10
and 40~arcsec) returns a value of $\Omega_m = 1.06 \pm ^{0.51}_{0.61}$
and for the NFW $\Omega_m \sim 1.3$, with large error.

The large error in Croom \& Shanks (1999) value for the space density
of groups remains a dominant systematic in our estimates of $\Omega_m$
and needs to be reassessed when group redshifts become available.
Some additional error may be introduced by a lack of accurate redshift
determinations for our galaxy groups.  Groups of galaxies that are actually
greatly separated in redshift may accidentally align and thus be
counted as a single halo, although it is unclear to what extent this
contamination could influence our lensing results, as any associations
along the line of sight still trace an increase
in the intervening mass distribution.  Certainly, associations
of galaxies that are separated greatly in redshift will not have dark
matter profiles like the SIS and NFW profiles used in our modelling,
being more like filaments than single haloes.  Ray tracing of high resolution, N-body 
simulations of the foreground mass distribution, where we can also
apply our group detection algorithms, are needed to test the size of the 
anti-correlation expected under a specific cosmological model such as 
$\Lambda$CDM.

Many of the inaccuracies in our method could be resolved by a large
sample of complementary galaxies 
and QSOs with redshift determinations and as such we await the publishing of
the 2dF Galaxy Redshift Survey \cite{Col01} in its entirety.
In the meantime, our estimates of $\Omega_m$ associated with groups continue to appear as high as 
those found by Croom \& Shanks.  However, the errors have increased meaning 
the rejection of $\Omega_m \sim 0.3$ is only at the 1-2$\sigma$ level and 
we have noted that there may be further systematic errors still to be 
taken into account. Nevertheless, it is still tantalising that a 
$\sigma=600~\rm{km~s^{-1}}$ dispersion is acceptable in at most 13/250 simulations
for groups as numerous as those used here and this clearly motivates 
the application of this technique  in larger QSO and galaxy group 
surveys.

\begin{figure} 
\begin{centering}
\begin{tabular}{c}
{\epsfxsize=8truecm \epsfysize=8truecm \epsfbox[55 170 540 705]{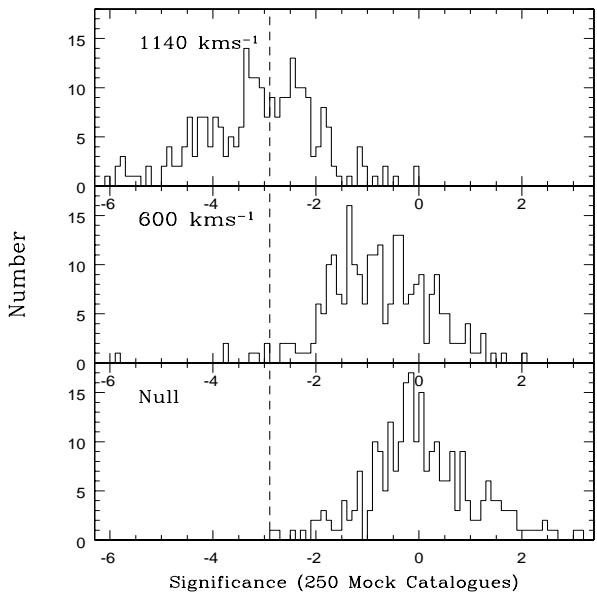}}
\end{tabular}
\caption
{The distribution of the strength of correlation for cross-correlations
  between 250 mock QSO catalogues and objectively derived galaxy groups.
  The mock QSO catalogues have been weighted to reflect a number of
  different possible SIS halo models.  The significance of the
  cross-correlation is measured out to 10~arcmin for each mock 
  QSO vs galaxy group cross-correlation function.  Cross-correlating
  2QZ QSOs against galaxy groups resulted in an anti-correlation of
  $-2.9\sigma$, which is marked as the dashed line.}
\label{fig:gauss.ps}
\end{centering}
\end{figure}

\subsection{Discussion}

The approach to the modelling of QSO lensing adopted in this paper is
different to that of some other authors.  Williams \& Irwin (1998)
used the galaxy-QSO cross-correlation function coupled with the
galaxy-galaxy auto-correlation function to derive constraints on
$\Omega_M/b$, where $b$ is the galaxy bias parameter and found their
phenomenological model largely agreed with sophisticated analytic
approaches (Dolag \& Bartelmann 1997; Sanz, Martinez-Gonzalez \&
Benitez 1997).  In a similar study utilising the cluster-QSO cross-correlation function coupled with 
the cluster-cluster auto-correlation, Croom \& Shanks (1999) obtained
an estimate of $\Omega/b \sim 3-4$.  However, this estimate
assumes that the total contribution to the $\omega_{cq}$ anti-correlation comes from
lensing by other clusters at a distance from the considered central cluster and
that the contribution from the central cluster is negligible, which may
not be true on scales a few arcminutes away from the
cluster centre.  The same criticism does not apply to
the Williams and Irwin (1998) result, it being detected on scales of
tens of arcminutes, where galaxy-galaxy clustering will dominate $\omega_{gq}$ rather
than the central galaxy halo. Their estimate of $\Omega_m/b \sim$
4-5 therefore stands, although their statistical errors may still allow a value of
$\Omega_m/b \sim$ 1.

Above, we determine the anti-correlation between galaxy groups and QSOs
is best fit by a high-mass NFW model ( $\sim 1.2 \times
10^{15}~h^{-1}~\msun$) or a high-velocity dispersion  SIS model ($\sim 1150~\rm{km~s^{-1}}$).
Although the preference for the $1140~\rm{km~s^{-1}}$ SIS group velocity dispersion
is only at the $\sim 2\sigma$ level in the current data, taking this result
together with the previous independently derived result of BFS88 and also
with the strong positive correlations seen in the brighter LBQS sample, it
is clearly worth considering the implications if the amplitude of the
anti-correlation was correct and caused wholly by weak lensing.
Dynamical analysis of 2dF galaxy z-space distortions, results in an
estimate of $\beta= \Omega_m^{0.6}/b \sim
0.43 \pm 0.07$ \cite{Pea01} and so measurement of $\Omega_m \sim 1$
would imply $b \sim 2.5$.  Although there is no immediate contradiction
with the result presented a contradiction does arise when current
CMB constraints on the mass power spectrum are included. These suggest
that the galaxy power spectrum is approximately unbiased, implying
$\Omega_m \sim 0.3$, in contradiction at the $2\sigma$ level with our best-fit
result.

There are other constraints on the mass of galaxy groups which are in contradiction with our best-fit
velocity dispersion. In particular, Hoekstra et al. (2001) have used shear to
measure weak lensing of galaxies behind CNOC groups, finding
$\sigma \sim 300~\rm{km~s^{-1}}$ and $M/L \sim 200h
M_{\sun}/L_{\sun}$. These translate into a value of
$\Omega_m=0.19$. It is not clear why there is a difference between their
results and ours.  If the result of Hoekstra is correct, we would have to
appeal to moderate statistical fluctuations to explain our high
anti-correlation amplitude in an $\Omega_m=0.3$ model. Clearly shear
studies behind nearby galaxy groups which also have QSO
lensing data will be valuable.

It should also be noted that the number of galaxies detected in N~$\ge$~7
groups corresponds to only 7.1~per~cent of the total number of
galaxies. These exist, on average, within 2.5~arcmin of the cluster centre. As mentioned in Section 3, our
group centre against galaxy correlation function ($\omega_{cg}$) agrees with 
Stevenson, Fong and Shanks (1988). Integrating $\omega_{cg}$ to a
radius of $r<10'$, the extent of our anti-correlation signal, suggests that
the total number of galaxies associated with our groups is actually
$\sim 20$ per cent.
If we assume that the M/L ratio of galaxies in clusters is the same as
outside the groups then our estimate of $\Omega_m$ would rise by a factor
of $\sim$ 5. However, in the $\Lambda$CDM model, for example, the galaxy
distribution is expected to be anti-biased on small scales ie the M/L in
clusters is expected to be higher by as much as $\sim$ 2 \cite{Col99} and this introduces a
further uncertainty into our best estimate of $\Omega_m$.   This suggests that our best
estimate of $\Omega_m\sim1$ might be considered a weak lower limit and our
1-2$\sigma$ value of $\Omega_m\sim0.3$ a strong lower limit.  Any stronger
conclusions from the present dataset await a more detailed test in a
$\Lambda$CDM N-body simulation of the foreground mass distribution. 

\section{Conclusions}

We have sought the effects of weak gravitational lensing in correlations
between 2QZ QSOs and galaxy groups derived from the SDSS or from the Southern
APM.  We confirm that there is a distinct ($\sim 3\sigma$) 
anti-correlation between objectively determined galaxy
groups and QSOs \cite{Boy88}.  The anti-correlation is fit well by
supposing its cause is gravitational lensing through dark matter haloes, either
NFW haloes or Singular Isothermal Spheres, but requires more mass than
models with $\Omega_m=0.3$ would suggest, at the $\sim$ 1-2$\sigma$
level. Larger QSO samples could not 
only better distinguish the amplitudes of the anti-correlation predicted by 
cosmological models but may also be able to 
distinguish between different forms for dark matter halo profiles.

The observed anti-correlation does not appear to be caused by a selection 
effect due to the limited spacing of 2dF fibres.  We also rule out the idea that intervening dust
causes the signal \cite{Boy88}.  Our 95~per~cent upper limits on reddening in $b_j-r$
are $0.012$, which corresponds to $A_B<0.04$~mag assuming
the usual Galactic dust law. To explain the anti-correlation by dust would
need $A_B\approx 0.2$~mag (see BFS88); the dust hypothesis could then 
only be saved by appealing to either a smooth distribution of grey dust 
or a patchy distribution of heavily line of sight distributed dust around
galaxy groups.  It is also not straightforward to rule out the
hypothesis that \textit{both} dust and lensing
play some combined role in producing the anti-correlation signal.

It seems that weak lensing remains the likely explanation for the anti-correlation
between QSOs and galaxy groups.  The strength of the anti-correlation
suggests that there may be more mass hidden in galaxy groups than many previous
estimates require, making further study of this phenomenon
worthwhile. The models used in Section 5.2 to describe the form of the correlation function
remain simplistic averages and it would be worth performing some large,
high-resolution simulations of the expected lensing influence of
foreground mass on QSO distributions in different cosmologies.
Accurate ray tracing through N-body simulations
would predict the expected anti-correlation for different cosmologies.
Additionally, it is probable that a large proportion of our groups are really 
chance alignments of galaxies that are actually greatly separated in redshift.  Running our group
detection procedure using galaxies from  a large simulation, should
allow us to determine how many of our
groups are genuine haloes and how many are associations or filamentary in structure.
This, in turn, would allow us to produce a more accurate model of the
mass profile we are fitting to the data.

On the observational side, new faint QSO surveys linked to the 2QZ should definitively
determine the faint-end slope of the deep QSO number-magnitude relation.
Redshift information in the 2dF Galaxy Redshift Survey should improve
our estimate of $\Omega_m$ a great deal.  Using forthcoming catalogues
of 2dFGRS galaxy groups (Eke et al., in preparation) it should be possible to accurately
look at the anti-correlation amplitude as a function of group/cluster richness to
try and further distinguish the masses associated with
groups and clusters.  Groups determined from galaxies of known redshift
have the added advantage of tracing definite dark matter haloes, rather 
than filaments or allignments.  Certainly, more statistical
power is also needed, motivating extending the 2dF QSO over a wider area
and to greater depth.  The large area
of the Sloan Digital Sky Survey, containing both QSO and galaxy
samples, may be useful for both accurate grouping of galaxies in
redshift space and measuring the extent of the anti-correlation on small scales, where the 
mass and its form is best constrained.  Notably, the SDSS will contain significant samples of QSOs
brighter than $b_j \sim 19$, allowing us to look for confirmation of the
expected positive correlation between galaxies and QSOs brighter than
the knee of the QSO number counts.

\section*{Acknowledgements}

ADM ackowledges the generous support of a PPARC studentship.  This
paper was prepared utilising the Starlink node at Durham University.

\appendix

\section{Statistical lensing of QSOs through dark matter haloes}

\subsection{General lensing}

As Einstein noted \cite{Ein15}, a mass $M$ will bend a ray of light passing at
impact parameter $b$ through an angle $\alpha$ \cite{Sch92}

\begin{eqnarray}
\alpha = \frac{4GM(<b)}{bc^2} = \frac{D_{\rm s}}{ \it D_{\rm{ls}}}(\it\theta - \theta_{\rm q})
\end{eqnarray}
where $D_{\rm l}$,$D_{\rm s}$ and $D_{\rm{ls}}$ are the angular diameter distances of the
lens as measured by the observer, the source as measured by the observer and
the source as measured by the lens, respectively; $\theta$ is the angle 
between the image, the observer and the centre of the lens; $\theta_{\rm q}$ is the angle between 
the source, the observer and the centre of the lens.  Flux conservation implies each
image will be magnified by a factor

\begin{eqnarray}
  \mu = |\frac{\theta}{\theta_{\rm q}}\frac{\rm{d}\theta}{\rm{d}\theta_q}|  
\end{eqnarray}
\cite{Tur84}.  Lensing influences a sample of background objects in two 
competing ways.  Fainter objects are lensed into a magnitude limitied
sample, increasing the number density of that sample but the area
behind the lens is proportionately expanded, reducing the sample's
number density.  Narayan (1989) quantified this effect as a `net enhancement factor'

\begin{eqnarray}
  q = \frac{1}{\mu}\frac{N(<m+2.5log(\mu))}{N(<m)}  
\end{eqnarray}
The Number-magnitude relation can be approximated as a power law,
with $N(<m)\propto10^{\beta m}$ \cite{Boy88}, allowing us to express the enhancement factor as

\begin{eqnarray}
  q = \frac{1}{\mu}\frac{10^{\beta(m+2.5log(\mu))}}{10^{\beta m}} =
  \mu^{2.5\beta - 1}  
\end{eqnarray}
\cite{Cro97}.  Now, the net enhancement factor is the ratio of the
observed (lensed) flux and the true (unlensed) flux.  The correlation
function $\omega(\theta)$ may be expressed as the ratio of observed pairs of objects to 
expected pairs of objects.  Typically, 1 is subtracted from the
correlation function to account for the expected normal background of
pairs.  Hence:

\begin{eqnarray}
  \omega(\theta) = q - 1 = \mu^{2.5\beta - 1} - 1
\end{eqnarray}

Equation (A5) dictates $\omega(\theta) = 0$ when $\beta = 0.4$.  For
higher values of $\beta$ we would observe a correlation, and for lower
values, an anti-correlation.  Thus, the lensing effect is dependent on
the slope of the number-magnitude relation.

\subsection{Dark matter profiles}

It is simple to integrate a Singular Isothermal Sphere (SIS) profile out to an 
impact parameter $b$ \cite{Cro97} to determine that it will magnify
background sources by a factor

\begin{eqnarray}
  \mu = |\frac{\theta}{\theta -4\rm{\pi}\frac{D_{\rm{ls}}}{D_{\rm s}}(\frac{\sigma}{c})^2}|
\end{eqnarray}

with $c$ being the speed of light and $\sigma$ the velocity dispersion
of the SIS.  The term
$4\rm{\pi}\frac{D_{\rm{ls}}}{D_{\rm{s}}}(\frac{\sigma}{c})^2$ in the denominator is the
Einstein Radius.  The Einstein Radius essentially marks the boundary
between the weak and strong lensing regimes.  Equations (A5) and (A6) can be combined to 
constrain the predicted form of the correlation function from magnification through an SIS profile.

N-body simulations have provided a universal (NFW) density profile for dark
matter haloes (Navarro, Frenk \& White 1995, 1996, 1997) that has been
independently observationally confirmed with some success (Bartelmann
et al. 1998; Thomas et al. 1998).  We have also constructed a simple model to determine the
form of the correlation function based on lensing through such a halo.

The NFW profile in the form

\begin{eqnarray}
  \rho(r) = \frac{\delta_{\rm{c}}\rho_{\rm{c}}}{\frac{r}{r_{\rm{s}}}(1+\frac{r}{r_{\rm{s}}})^2} 
\end{eqnarray}

where $r_{\rm{s}}$ is a representative radial scale and $\rho_{\rm{c}}$ is the
critical density or the universe at the redshift of the dark matter, appears to be a reasonable
description of haloes spanning 9 orders of magnitude in mass, from
globular clusters to galaxy clusters.  Ideally, we would wish to study
statistical lensing utilising just such a realistic density profile.
Following Bartelmann (1996), we rewrite the profile as

\begin{eqnarray}
  \rho(x) = \frac{\rho_{\rm{s}}}{x(1+x)^2} 
\end{eqnarray}

and consider lensing around the profile at a radius of the impact
parameter $b$, meaning that $x$ is defined by

\begin{eqnarray}
  x \equiv \frac{b}{r_{\rm{s}}} \equiv \frac{D_{\rm{l}}\theta}{r_{\rm{s}}}
\end{eqnarray}

where $D_{\rm{l}}$ and $\theta$ are defined as for the SIS, above.  One can then combine
Equations (A2) and (A9) to yield:

\begin{eqnarray}
   \mu = |(\frac{r_{\rm{s}}}{D_{\rm{l}}})^2\frac{x}{\theta_{\rm{q}}}\frac{{\rm{d}}x}{{\rm{d}}\theta_{\rm{q}}}|
\end{eqnarray}

Now, Maoz et al. (1997) determine the characteristic scale $r_{\rm{s}}$ from
an empirical fit to Figure 9 of Navarro et al. (1997) as

\begin{eqnarray}
  r_{\rm{s}} = 300(\frac{M}{10^{15} \msun})^{\gamma}& \kpc
\end{eqnarray}

where $h$ is the Hubble constant in units of 100$~\rm{km~s^{-1} Mpc^{-1}}$ and
$\gamma$ varies between cosmogonies (for CDM, $\gamma \sim \frac{1}{3}$).
Maoz et al. (1997) go on to rewrite Equation (A1) as

\begin{eqnarray}
  \alpha = \frac{4GM_{1.5}}{c^2r_{\rm{s}}g(1.5Mpc/r_{\rm{s}})}\frac{g(x)}{x} \\
\end{eqnarray}

where $M_{1.5}$ is the mass within $1.5\mpc$ of the centre of the halo and
$g(x)$ is given by Bartelmann (1996) as:

\begin{eqnarray}
  g(x) = \ln\frac{x}{2}+\frac{2}{x^2-1}\tan^{-1}\sqrt{\frac{x-1}{x+1}}
   & (x>1) \\
  g(x) = \ln\frac{x}{2}+\frac{2}{1-x^2}\tanh^{-1}\sqrt{\frac{1-x}{1+x}}
   & (x<1) \\
  g(x) = \ln\frac{x}{2}+1  &  (x=1)
\end{eqnarray}

Combining Equations (A1) and (A12), we can deduce

\begin{eqnarray}
  \theta_{\rm{q}} = \theta-\frac{D_{\rm{ls}}}{D_{\rm{s}}}\frac{4GM_{1.5}}{r_{\rm{s}}c^2g(1.5Mpc/r_{\rm{s}})}\frac{g(x)}{x} \\
  \frac{{\rm{d}}\theta_{\rm{q}}}{{\rm{d}}x} = \frac{{\rm{d}}\theta}{{\rm{d}}x}-\frac{D_{\rm{ls}}}{D_{\rm{s}}}\frac{4GM_{1.5}}{r_{\rm{s}}c^2g(1.5Mpc/r_{\rm{s}})}\frac{{\rm{d}}\frac{g(x)}{x}}{{\rm{d}}x}
\end{eqnarray}

and substitute in Equation (A9) to ultimately find

\begin{eqnarray}
  \frac{\theta_{\rm{q}}}{x} = \frac{r_{\rm{s}}}{D_{\rm{l}}}-\frac{D_{\rm{ls}}}{D_{\rm{s}}}\frac{4GM_{1.5}}{r_{\rm{s}}c^2g(1.5Mpc/r_{\rm{s}})}\frac{g(x)}{x^2}\\
  \frac{{\rm{d}}\theta_{\rm{q}}}{{\rm{d}}x} = \frac{r_{\rm{s}}}{D_{\rm{l}}}-\frac{D_{\rm{ls}}}{D_{\rm{s}}}\frac{4GM_{1.5}}{r_{\rm{s}}c^2g(1.5Mpc/r_{\rm{s}})}\frac{{\rm{d}}\frac{g(x)}{x}}{{\rm{d}}x}
\end{eqnarray}

which can be readily substituted into Equation (A10) and numerically
solved to derive the magnification.  The magnification dictates the
expected form of the correlation function as outlined in Equation (A5).

\subsection{Cosmology: Determining $\Omega_m$ in Groups}

We will now briefly outline how the mass density of the Universe might
be determined from our study, noting that the value thus derived is
purely a measure of $\Omega_m$ in galaxy groups.  $\Omega_m$ may be defined

\begin{eqnarray}
  \Omega_m = \frac{\rho_0}{\rho_{\rm{crit}}} = \frac{8\rm{\pi}G}{3H_0^2}\rho_0
\end{eqnarray}
where $\rho_0$ can be estimated as the product of the space density of
galaxy groups and 
the average mass of a dark matter halo.  The sky density of galaxy
groups can be measured.  Equations (A5) and (A6) relate the average velocity
dispersion of the SIS to the observed correlation function.  The
average mass of the SIS may be derived from the average velocity
dispersion, yielding an estimate of $\Omega_m$ in groups of

\begin{eqnarray}
  \Omega_m = \frac{8\rm{\pi} G}{3H_0^2}nM_{\rm{SIS}} =  \frac{8\rm{\pi}^2 }{3H_0^2}n\sigma^2 r
\end{eqnarray}
where n is the space density of galaxy groups and r refers to the
extent of the dark matter halo or, roughly, the extent of
correlation. Similarly, we can express the
correlation function in terms of the mass within  $1.5\mpc$ of the
centre of an NFW halo.  Theoretically, then, $\Omega_m$ is alternatively given by

\begin{eqnarray}
  \Omega_m = \frac{8\rm{\pi} G}{3H_0^2}nM_{\rm{NFW}} \geq \frac{8\rm{\pi} G}{3H_0^2}nM_{1.5}
\end{eqnarray}

where $M_{1.5}$ is the mass within $1.5\mpc$ of the centre of an NFW halo.


\begin{thebibliography}{}

\bibitem[\protect\citename{Abell }1958]{Abe58} Abell~G.~O., 1958, ApJS, 3, 211
\bibitem[\protect\citename{Abell, Corwin \& Olowin }1989]{Abe89} Abell~G.~O., Corwin~H.~G., Olowin~R.~P., 1989, ApJS, 70, 1
\bibitem[\protect\citename{Bartelmann }1996]{Bar96} Bartelmann~M., 1996, A\&A, 313, 697
\bibitem[\protect\citename{Bartelmann et al.  }1998]{Bar98} Bartelmann~M., Huss~A.,
Colberg~J.~M., Jenkins~A., Pearce~F~R., 1998, A\&A, 330, 1 
\bibitem[\protect\citename{Baugh \& Efstathiou }1993]{Bau93} Baugh~C.~M., Efstathiou~G., 1993, MNRAS, 265, 145
\bibitem[\protect\citename{Benitez, Sanz \& Martínez-González }2001]{Ben01} Benitez~N., Sanz~J.~L., Martinez-Gonzalez~E., 2001, MNRAS, 320, 241
\bibitem[\protect\citename{Boyle, Fong \& Shanks }1988]{Boy88} Boyle~B.~J., Fong~R., Shanks~T., 1988, MNRAS, 231, 897
\bibitem[\protect\citename{Boyle, Jones \& Shanks }1991]{Boy91} Boyle~B.~J.,
Jones~L.~R., Shanks~T., 1991, MNRAS, 251, 482
\bibitem[\protect\citename{Boyle et al. }2000]{Boy00} Boyle~B.~J., Shanks~T., Croom~S.~M., Smith~R.~J., Miller~L., Loaring~N., Heymans~C., 2000, MNRAS, 317, 1014
\bibitem[\protect\citename{Broadhurst, Taylor \& Peacock }1995]{Bro95} Broadhurst~T.~J., Taylor~A.~N., Peacock~J.~A., 1995, ApJ, 438, 49
\bibitem[\protect\citename{Bukhmastova }2001]{Buk01} Bukhmastova~Yu.~L., 2001,  ARep, 45, 581
\bibitem[\protect\citename{Colless }1998]{Col98} Colless~M., 1998, in `Wide Field Surveys in Cosmology', Editions Frontieres, ISBN 2-8 6332-241-9, 77
\bibitem[\protect\citename{Colless }2001]{Col01} Colless~M., et al., 2001, MNRAS, 328, 1039
\bibitem[\protect\citename{Colin et al. }1999]{Col99} Colin~P., Klypin~A.~A., Kravtsov~A.~V., Khokhlov~A.~M., 1999, ApJ, 523, 32
\bibitem[\protect\citename{Croom }1997]{Cro97} Croom~S.~M., 1997, PhD Thesis, University of Durham
\bibitem[\protect\citename{Croom \& Shanks }1999]{Cro99}Croom~S.~M., Shanks~T., 1999, MNRAS, 307, 17
\bibitem[\protect\citename{Croom et al. }2001]{Cro01} Croom~S.~M., Smith~R.~J., Boyle~B.~J., Shanks~T., Loaring~N~S., Miller~L., Lewis~I~J.,  2001, MNRAS, 322, 29
\bibitem[\protect\citename{Dalton et al. }1997]{Dal97} Dalton~G.~B., Maddox~S.~J., Sutherland~W.~J., Efstathiou~G., 1997, MNRAS, 289, 263
\bibitem[\protect\citename{Dolag \& Bartelmann }1997]{Dol97} Dolag~K., Bartelmann~M.,  1997, MNRAS, 291, 446
\bibitem[\protect\citename{Einstein }1915]{Ein15} Einstein~A., `Erkl\"arung der
Perihelbewegung des Merkur aus der allgemeinen Relitivit\"atstheorie',Sitzunberger. Preu\ss. Akad. Wissench., erster Halbband, p. 831
\bibitem[\protect\citename{Ferguson }1993]{Fer93} Ferguson~H.~C., 1993, MNRAS, 263, 343
\bibitem[\protect\citename{Ferreras et al. }1997]{Fer97} Ferreras~I., Benitez~N., Martinez-Gonzalez~E., 1997, AJ, 114, 1728
\bibitem[\protect\citename{Girardi et al. }1992]{Gir92} Girardi~M., Mezzetti~M., Giurcin~G., Mardirossian~F., 1992, ApJ, 394, 442 
\bibitem[\protect\citename{Gaztanaga }2002]{Gaz02} Gaztanaga~E., astro-ph/0210311
\bibitem[\protect\citename{Hartwick \& Schade }1990]{Har90} Hartwick~F.~D.~A., Schade~D., 1990, ARA\&A, 28, 437
\bibitem[\protect\citename{Hoekstra et al. }2001]{Hoe01} Hoekstra~H., Franx~M., Kuijken~K., Carlberg~R.~G., Yee~H.~K.~C., Lin~H., Morris~S.~L., Hall~P.~B., Patton~D.~R., Sawicki~M., Wirth~G.~D., 2001, ApJ, 548, 5
\bibitem[\protect\citename{Koo \& Kron }1988]{Koo88} Koo~D.~C., Kron~R.~G., 1988,
ApJ, 325, 92 
\bibitem[\protect\citename{Maddox et al. }1990a]{Mad90a} Maddox~S.~J., Efstathiou~G.,
Sutherland~W.~J., Loveday~ J., 1990, MNRAS, 243, 692
\bibitem[\protect\citename{Maddox et al. }1990b]{Mad90b} Maddox~S.~J.,
  Efstathiou~G., Sutherland~W.~J., 1990, MNRAS, 246, 433
\bibitem[\protect\citename{Maoz }1995]{Mao95} Maoz~D., 1995, ApJ, 455, 115
\bibitem[\protect\citename{Maoz et al. }1997]{Mao97} Maoz~D., Rix~H.~W., Gal-Yam~A., Gould~A., 1997, AJ, 486, 75
\bibitem[\protect\citename{Martinez et al. }1999]{Mar99} Martinez~H.~J., Merchan~M.~E., Valotto~C.~A., Lambas~D.~G., 1999, ApJ, 514, 558
\bibitem[\protect\citename{Menard \& Bartelmann }2002]{Men02} Menard~B., Bartelmann~M.,  2002, A\&A, 386, 784
\bibitem[\protect\citename{Miller et al. }2003]{Mil03} Miller~L., Lopes~A.~M., Smith~R.~J., Croom~S., Boyle~B.~J., Shanks~T., Outram~P., astro-ph/0210644  
\bibitem[\protect\citename{Narayan }1989]{Nar89} Narayan~R., 1989, ApJ, 339, 53
\bibitem[\protect\citename{Navarro, Frenk \& White }1995]{Nav95} Navarro~J.~F.,
Frenk~C.~S., White~S.~D.~M., 1995, ApJ, 275, 720
\bibitem[\protect\citename{Navarro, Frenk \& White }1996]{Nav96} Navarro~J.~F.,
Frenk~C.~S., White~S.~D.~M., 1996, ApJ, 462, 563
\bibitem[\protect\citename{Navarro, Frenk \& White }1997]{Nav97} Navarro~J.~F.,
Frenk~C.~S., White~S.~D.~M., 1997, AJ, 486, 75
\bibitem[\protect\citename{Norman \& Impey }2001]{Nor01} Norman~D.~J., Impey C.~D., 2001, AJ, 121, 2392
\bibitem[\protect\citename{Norman \& Williams }2000]{Nor00} Norman~D., Williams~L.~L.~R., 2000, AJ, 119, 2060
\bibitem[\protect\citename{Peacock et al. }2001]{Pea01} Peacock~J.~A., et al., 2001, Nature, 410, 169
\bibitem[\protect\citename{Peebles }1980]{Pee80} Peebles~P.~J.~E., 1980, in `The Large Scale Structure in the Universe', Princeton University Press, ISBN 0-691-08240-5 
\bibitem[\protect\citename{Rodrigues Williams \& Hogan}1994]{Rod94} Rodrigues-Williams~L.~L., Hogan~C., 1994, AJ, 107, 451
\bibitem[\protect\citename{Sanz, Martinez-Gonzalez \& Benitez }1997]{San97} Sanz~J.~L., Martinez-Gonzalez~E,  Benitez~N., 1997, MNRAS, 291, 418
\bibitem[\protect\citename{Schlegel, Finkbeiner \& Davis }1998]{Sch98} Schlegel~D.~J.,
Finkbeiner~D.~P., Davis~M., 1998, ApJ, 500, 525
\bibitem[\protect\citename{Schneider, Ehlers \& Falco }1992]{Sch92} Schneider~P.,
Ehlers~J., Falco~E.~E., 1992, in `Gravitational Lenses', Springer-Verlag, ISBN 0-387-97070-3, Section 2
\bibitem[\protect\citename{Shanks et al. }1983]{Sha83} Shanks~T., Fong~R., Green~M.~R., Clowes~R.~G., Savage~A., 1983, MNRAS, 203, 181
\bibitem[\protect\citename{Smith et al. }1997]{Smi97} Smith~R.~J., Boyle~B.~J., Shanks~T., Croom~S.~M., Miller~L., Read~M., 1997, IAUS, 179, 348
\bibitem[\protect\citename{Squires \& Kaiser }1996]{Squ96} Squires~G., Kaiser~N., 1996, ApJ, 473, 65 
\bibitem[\protect\citename{Stevenson, Fong \& Shanks }1988]{Ste88}
Stevenson~P.~R.~F., Fong~R., Shanks~T., 1998, MNRAS, 234, 801
\bibitem[\protect\citename{Stoughton et al. }2002]{Sto02} Stoughton~C. et al., 2002, AJ, 123, 485
\bibitem[\protect\citename{Thomas et al. }1998]{Tho98} Thomas~P.~A., Colberg~J.~M.,
Couchman~H.~M.~P., Efstathiou~G.~P., Frenk~C.~S., Jenkins~A.~R.,
Nelson~A.~H., Hutchings~R.~M., Peacock~J.~A., Pearce~F.~R., White~S.~D.~M., 1998, MNRAS, 296, 1061
\bibitem[\protect\citename{Turner \& Gott }1976]{Tur76} Turner~E.~L., Gott III~J.~R., 1976, ApJS, 32, 409
\bibitem[\protect\citename{Turner, Ostriker \& Gott }1984]{Tur84} Turner~E.~L., Ostriker~J.~P., Gott III~J.~R., 1984, ApJ, 284, 1
\bibitem[\protect\citename{Wambsganss }1998]{Wam98} Wambsganss~J., 1998, `Gravitational Lensing in Astronomy', Living Rev. Relativity 2, 1-80, http://www.livingreviews.org/ Articles/Volume1/1998-12wamb/ 
\bibitem[\protect\citename{Williams \& Irwin }1998]{Wil98} Williams~L.~L.~R., Irwin~M., 1998, MNRAS, 298, 378
\bibitem[\protect\citename{Wu }1994]{Wu94} Wu~X.~P., 1994, A\&A, 286, 748
\bibitem[\protect\citename{Wu et al. }1996]{Wu96} Wu~X.~P., Fang~L.~Z., Zhu~Z.~H., Qin~B., 1996, ApJ, 307, 575
\bibitem[\protect\citename{Yasuda et al. }2001]{Yas01} Yasuda~N., et al., 2001, AJ, 122, 1104
\bibitem[\protect\citename{York et al. }2000]{Yor00} York~D.~G., et al., 2000, AJ, 120, 1579

\end{thebibliography}
\end{document}